\documentclass[twocolumn]{aastex631}
\pdfoutput=1
\usepackage{lineno}
\submitjournal{The Astronomical Journal}

\received{}
\revised{}
\accepted{}

\shorttitle{Galactic disk morphology}
\shortauthors{Joshi et al. (2023)}

\begin{document}
\title{Revisiting Galactic disk and spiral arms using open clusters}
\author[0000-0001-8657-1573]{Yogesh C. Joshi}
\correspondingauthor{Yogesh C. Joshi}
\email{yogesh@aries.res.in}
\affiliation{Aryabhatta Research Institute of Observational Sciences (ARIES), Manora Peak, Nainital 263002, India}
\author[0000-0012-3245-1234]{Sagar Malhotra}
\affiliation{Indian Institute of Science Education \& Research, Mohali 140306, India}

\begin{abstract}
We use the largest open clusters catalogue in the post-Gaia era to provide the observational view of the Galactic disk. By compiling the physical parameters like age, distance, and kinematic information, we investigate the spatial distribution of the open clusters and revisit the spiral arms and other asymmetries in the Galactic disk. Using the young open clusters as a tracer of spiral arms, we map the spiral structure of the Galaxy and found that most of the clusters start migrating away from the spiral arms in about 10-20 Myr and fill the inter-arm regions as their age progress. Using the 3D kinematic information of 371 open star clusters, we derived different individual pattern speeds for spiral arms closely following the Milky Way rotation curve, hence favouring the transient nature of Milky Way spiral arms. The pattern rotation speeds of each spiral arm suggest that the spiral arms have not accelerated in the last 80 Myrs. Based on the distribution of open clusters younger than 700 Myr above or below the Galactic plane, we found a Solar offset of \(z_\odot = 17.0\pm0.9\) pc north of the Galactic plane and estimated the scale height $z_h = 91.7 \pm 1.9$ pc from the Galactic plane.
\end{abstract}
\section{Introduction}\label{sec:intro}
Open star clusters (OCs) are a group of coeval stars that are formed as a clump in giant molecular clouds and provide vital  clues about understanding star formation and evolution in the Galaxy \citep{2010ARA&A..48..431P, 2019ARA&A..57..227K}. As most OCs are relatively younger populations and mainly lie within the Galactic disk, these sources are being viewed as excellent markers to study the structural, kinematic, and chemical properties of the Galactic disk through analysis of a large number of clusters in the Galaxy. Many authors have carried out such studies in the past \citep[cf.,][]{2006A&A...446..121B, 2014MNRAS.444..290B, 2016A&A...593A.116J, 2020AA...640A...1C, 2022MNRAS.517..675D}. Moreover, a wide range of ages of OCs, from a few million years to nearly 10 billion years, provides a unique opportunity to study the evolution of the Galactic disk and trace the star formation history across the Milky Way. In recent years, the number of OCs has grown multi-fold \citep{universe8020111, 2022A&A...661A.118C, 2023ApJS..266...36C}, paving the way for a better understanding of the Galactic disk, thanks to large-scale surveys and machine-learning methods \citep{2018AA...618A..93C, 2018ApJ...856..152R, 2019JKAS...52..145S, 2019A&A...624A.126C, 2019ApJS..245...32L, 2021MNRAS.504..356D, 2021RAA....21...93H, 2022AA...660A...4H, 2022AA...661A.118C}. The accurate physical parameters for OCs are obtained through analysis of precise membership identification of cluster stars facilitated through their astrometric and kinematic study \citep{2014A&A...564A..79D, 2018AA...618A..93C, 2021MNRAS.504..356D, 2022A&A...661A.118C, 2022AJ....164...54Z}.
%
\begin{table*}[ht]
  \centering
  \caption{List of references used to compile the catalog of 6133 OCs. In some cases, we have considered only subset of given sample - High Quality Sample\(^{(a)}\); Reliable Proper Motions \(^{(b)}\).}
    \begin{tabular}{|cc|cc|}
    \hline
    \textbf{Reference} & \textbf{No. of OCs} &   \textbf{Reference} & \textbf{No. of OCs}\\
    \hline
    {\cite{2022AA...660A...4H}} & 704 & {\cite{2020AA...633A..99C}} & 1481 \\
    \hline
    {\cite{2022AA...668A..13H}} & 38  & {\cite{2020MNRAS.496.2021F}} & 59\\
    \hline
    {\cite{2022AA...661A.118C}} & 628 & {\cite{2019AA...623A.108B}} & 269  \\
    \hline
    {\cite{2022ApJS..260....8H}} & 541 &  {\cite{2019MNRAS.483.5508F}} & 4  \\
    \hline
    {\cite{2022ApJS..262....7H}} & 886  & {\cite{2019AA...623A..80C}} & 90\\
    \hline
    {\cite{2022AA...659A..59T}} & 389 & {\cite{2018AA...618A..93C}} & 1229 \\
    \hline
    {\cite{2022AJ....164...85M}} & 150 &   {\cite{2018AJ....156..142D}} & 19  \\
    \hline
    {\cite{2022MNRAS.509..421N}} & 136 & {\cite{2016AA...585A.150N}} & 172  \\
    \hline
    {\cite{2021MNRAS.504..356D}} & 1743 &  {\cite{2015AA...581A..39S}} & 63\\
    \hline
    {\cite{2021AA...647A..19T}} & 393\(^{(a)}\)  & {\cite{2014AA...568A..51S}} & 139 \\
    \hline
    {\cite{2021AA...652A.102H}} & 3794  & {\cite{2013AA...558A..53K}} & 3006 \\
    \hline
    {\cite{2021MNRAS.503.3279S}} & 226  & {\cite{2013MNRAS.436.1465B}} & 775 \\
    \hline
    {\cite{2021ApJ...912....5H}} & 265   & {\cite{2010MNRAS.407.2109V}} & 481\\
    \hline
    {\cite{2021AA...652A..25C}} & 47  & {\cite{2010AstL...36...75G}} & 194 \\
    \hline
    {\cite{2020AA...640A...1C}} & 2017   & {\cite{2008AA...485..303M}} & 166  \\
    \hline
    {\cite{2020AJ....159..199D}} & 128  & {\cite{2007AN....328..889K}} & 516 \\
    \hline
   {\cite{2020ApJ...903...55P}} & 34\(^{(a)}\)  & {\cite{2003AJ....125.1397C}} & 144 \\
    \hline
    {\cite{2020AJ....160..279K}} & 1910   & {\cite{2002AA...389..871D}} & 2167 \\
    \hline
   {\cite{2020AA...640A.127Z}} & 295  & {\cite{2003ARep...47....6L}} & 347\(^{(b)}\)\\
    \hline
    \end{tabular}%
  \label{tab:1}%
\end{table*}%

It is important to note that much of this progress has come from the Gaia survey data, which provide unprecedented homogeneous and precise photometry, high-precision proper motion, and parallaxes of more than a billion sources \citep{2016A&A...595A...2G, 2018A&A...616A...1G} thereby leading to more precise and accurate estimation of the astrometric and physical parameters of OCs as compared to some pre-Gaia catalogues such as \citet{2002AA...389..871D} and \citet{2013AA...558A..53K}. 

Although substantial progress has been made in the past decade to understand the Galactic disk, it is still unclear what mechanism exactly drives the spiral arm formation in the Galaxy leading to a lack of consensus among the community on the nature of the spiral arms. In the present study, we aim to use the largest open clusters catalogue compiled in the post-Gaia era to provide the observational view of the Galactic disk. This paper is organized as follows: we describe the data used in the present study in Section~\ref{data}. The spatial distribution of the clusters in the Galaxy and spiral arm morphology and kinematics are described in Section~\ref{spatial}. Section~\ref{cluster_distri} discusses the OC distribution along and perpendicular to the Galactic plane. The summary of our results is laid out in Section~\ref{summary}.
\section{Data}\label{data}
We compiled a large catalog of OCs from more than 35 previous studies providing their mean astrometric and physical parameters, including the age and distance. The list of references used in compiling the catalog, along with the corresponding number of OCs analysed in each study is provided in Table \ref{tab:1}. The priority order of the references while compiling the catalog was decided by their year of publication and the Gaia data release used for studying OCs. Furthermore, since several clusters detected in the pre-Gaia era were found to be false positives when re-investigated in the Gaia data \citep[e.g.;][]{2020AA...633A..99C}, we have only included those OCs that have been discovered or re-visited in the Gaia era.
%
\begin{figure*}
\centering
\includegraphics[scale=0.45]{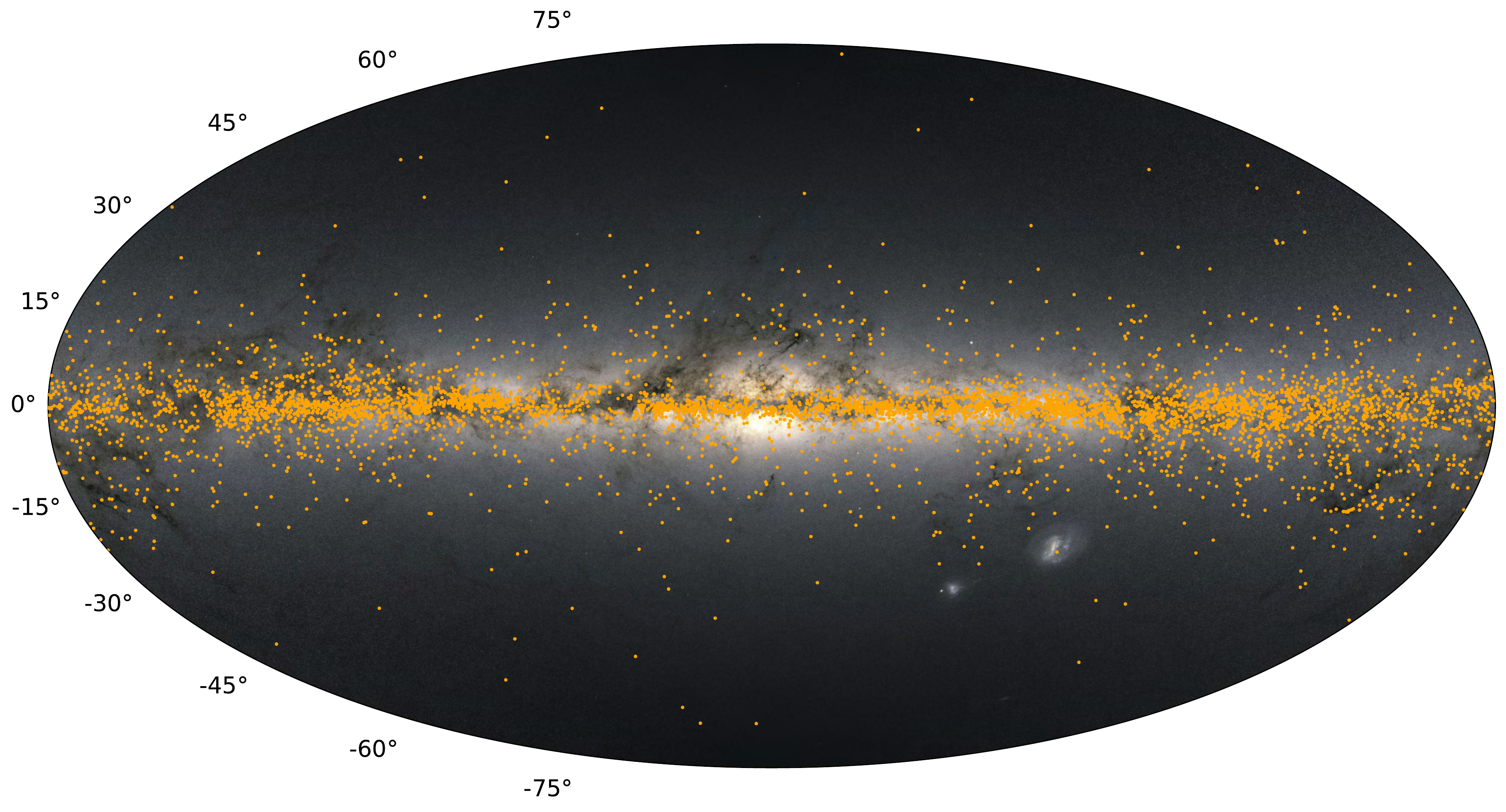}
\caption{An on-sky view of Galaxy generated using mw-plot in the Galactic longitude-latitude ($l-b$) plane juxtaposed with our sample of 6133 OCs. The image in the background is generated by ESA Gaia EDR3 with a resolution of 15.38 lightyears per pixel.}
\label{fig01}
\end{figure*}
%
While prioritizing the latest studies and giving lesser weight to those showing multiple discrepancies, we compiled a catalogue of 6133 OCs having information about positions, parallax, age, heliocentric distance, proper motions, and radial velocities, though not all the parameters are available for all the clusters. We preferred the cluster parameters from post-Gaia catalogs and only resorted to the pre-Gaia studies if no information was found in recent studies. Therefore, the final value of the parameter was the one provided in the most recent version of the catalog. Of the 6133 OCs in the catalogue, all have their age estimates available, however, only 4378 clusters have distance information. Among them, 2959 OCs have distance estimates from the Gaia data while the remaining 1419 OCs have distances retrieved from pre-Gaia era studies. Using the distance information, we estimated $(X,Y,Z)$ cartesian coordinates and the galactocentric distance $R_{GC}$ for all the 4378 OCs. In addition to this, the full 3D kinematic information is also available for 2483 OCs. Here, one has to note that since there are very few catalogs that provide the uncertainties in the estimated physical parameters of OCs, especially the distance, we have not used the distance uncertainties in our analysis.
\section{Spatial distribution}\label{spatial}

\subsection{2D distribution of OCs and present-day Spiral Arms}
The knowledge of the spatial distribution of OCs is important to understand the inhomogeneous nature of the Galaxy and probing the large-scale Galactic structure. For this purpose, we capitalized the positional coordinates of all the 6133 OCs to infer the broader picture of the structure of the Milky Way. In Figure~\ref{fig01}, we use mw-plot\footnote{https://pypi.org/project/mw-plot/} to show an on-sky view of the Galaxy in the Galactic longitude-latitude ($l-b$) plane with OCs juxtaposed in the $l-b$ plane wherein the image in the background is generated by ESA Gaia EDR3 and has a resolution of 15.38 light years per pixel. It is evident that there are few pockets where cluster density is low like \(50^{o}\) and \(150^{o}\) Galactic longitude regions. The lack of OCs in a region near $l \sim 150^{o}$ might be due to the lower density of gas and dust in the anti-center direction and has also been noticed earlier and dubbed as Gulf of Camelopardalis \citep{2019A&A...624A.126C, 2019A&A...627A..35C}. The deficiency of clusters around \(l \sim 50^{o}\) region might be understood by the higher disruption rate complemented by our obscured view in the direction of the Galactic centre. The distribution of clusters is more scattered around the Galactic mid-plane towards the Galactic anti-center direction than in the direction of the Galactic center. However, in general, most of the clusters are confined very close to the Galactic mid-plane, and number density sharply decreases as one goes away from the mid-plane with more than \(85\%\) of the total number of OCs belonging to [\(-10^{o}\), \(10^{o}\)] of the Galactic latitude while \(75\%\) of them belong to central [\(-5^{o}\), \(5^{o}\)] region.

\begin{figure}
\centering
\includegraphics[scale=0.27]{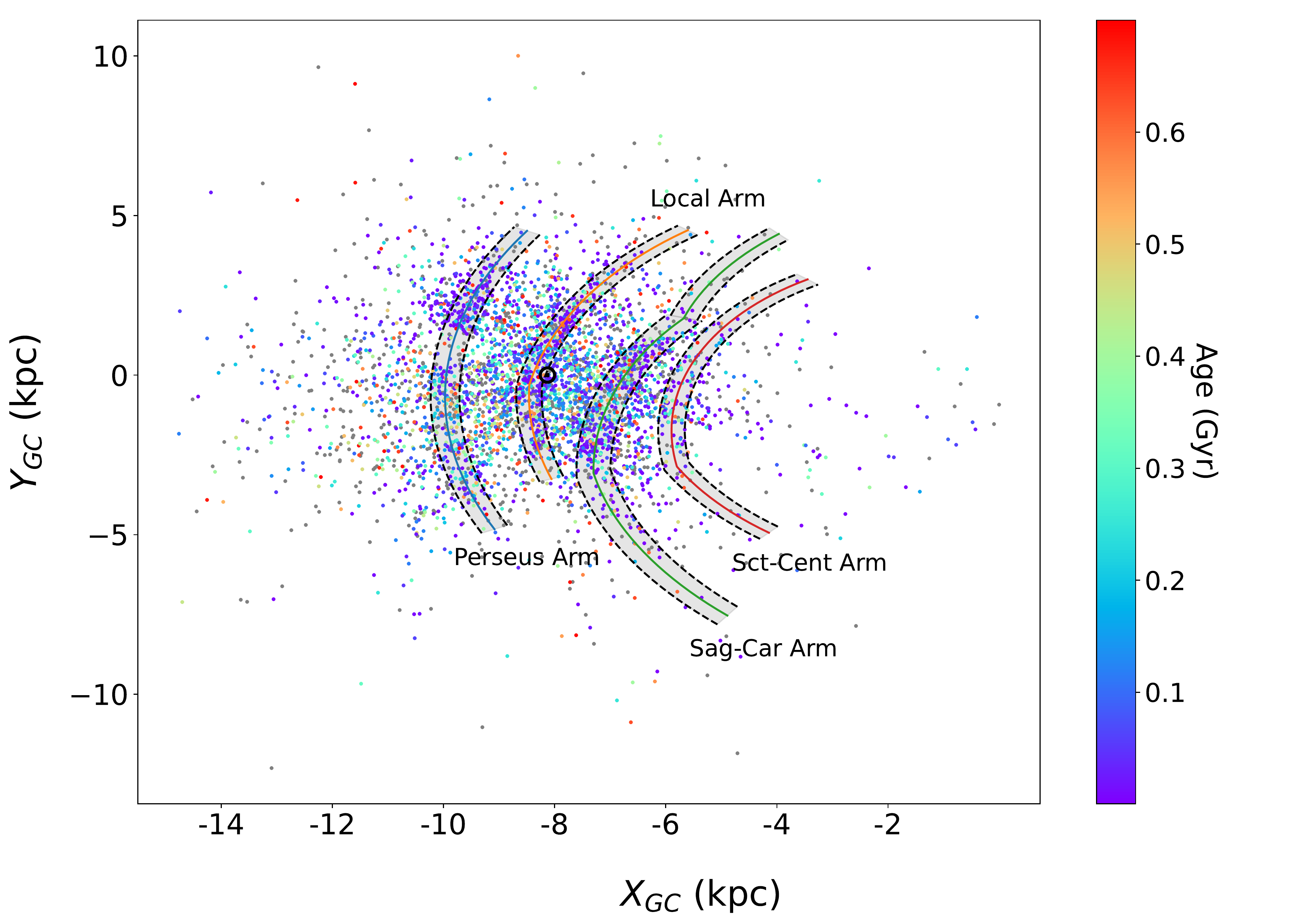}
\caption{Distribution of the complete sample of OCs in the $X-Y$ plane with the Galactic center at the origin. OCs younger than 700 Myr are color-coded based on their age, while the older ones are shown in gray color. The position of the Sun is marked with a black circle at (-8.15, 0) kpc. Spiral Arms are taken from \citet{2021FrASS...8..103H}.}
\label{fig02}
\end{figure}

To better resolve the multiple structures at various distances which are otherwise superposed on the sky, we illustrate the $X-Y$ distribution of OCs in the left panel of Figure~\ref{fig02} having the Sun at (-8.15,0) kpc \citep{2019ApJ...885..131R} and Galactic center at the origin. Here, we draw only 4378 OCs for which distance information was available. The OCs detected far from the Galactic center are not shown in this plot for better visualization. The OCs older than 700 Myr are shown in gray color while the younger ones are colored according to their ages.
%
%
We overplotted the usual log-periodic spiral arms taken from \citet{2021FrASS...8..103H} where the profile of each arm is described by \citet{2019ApJ...885..131R} as follows:
$$
\ln (R/R_{kink}) = -(\beta - \beta_{kink})\tan \psi
$$
where $R$ is the Galactic radius at a galactocentric azimuth angle \(\beta\) which is 0 along the line joining the Sun and the Galactic center and increases in the direction of Galactic rotation. \(R_{kink}\) and \(\beta_{kink}\) are the corresponding values of $R$ and \(\beta\) at the kink position, where there might be an abrupt change in the pitch angle \(\psi\). There has been substantial evidence in the recent past that spiral arms in our galaxy as well as other galaxies are made up of small segments having a scale length of about
5-8 kpc which are separated by kinks/gaps \citep{2015ApJ...800...53H}. A similar observation was made in N body simulations by \citet{2013ApJ...766...34D}.  Recent studies by \citet{2019ApJ...885..131R} and \citet{2021FrASS...8..103H} use accurate VLBI measurements of parallaxes of masers, etc. to fit log-periodic spiral arms while allowing for ‘kinks’ resulting in improved quality of fits. This model of allowing kinks in the spiral arms is supported by several studies \citep[e.g.;][]{2011MNRAS.414..538K, 2015ApJ...800...53H, 2019A&A...631A..94D} and found that spiral arms in different galaxies were not perfectly described by log-periodic spirals but rather were separated by gaps/kinks. Similarly, other studies \citep[e.g.;][]{1993ApJ...411..674T, 2014A&A...569A.125H} on Milky Way spiral arms noticed signatures of kinked spiral arms, segments of which could be fitted with log-periodic spiral functions with different pitch angles. Following such observations, we do not restrict our analysis to perfectly described log-periodic spirals but use the ones that allow for the existence of gaps in the observed spiral arms. \citet{2021FrASS...8..103H} models each of the spiral arms with one kink except the Sagittarius-Carina Arm which is described using two kinks at \(\beta\) = \(-22.8^\circ\) and \(17.5^\circ\), respectively. In Figure~\ref{fig02}, the regions enclosed within the dashed lines around the spiral arms correspond to the arm widths. The distribution of OCs in spiral arms is not uniform but shows a patchy appearance. It is observed that younger clusters populate the spiral arms while older clusters are mainly present in the inter-arm regions as has also been reported in previous studies by \citet{2020AA...633A..99C, 2021FrASS...8...62M, 2021FrASS...8..103H, 2021AA...647A..19T}, and \citet{2021A&A...651A.104P}.

\begin{figure}
\centering
\includegraphics[width=9.5cm,height=17.5cm]{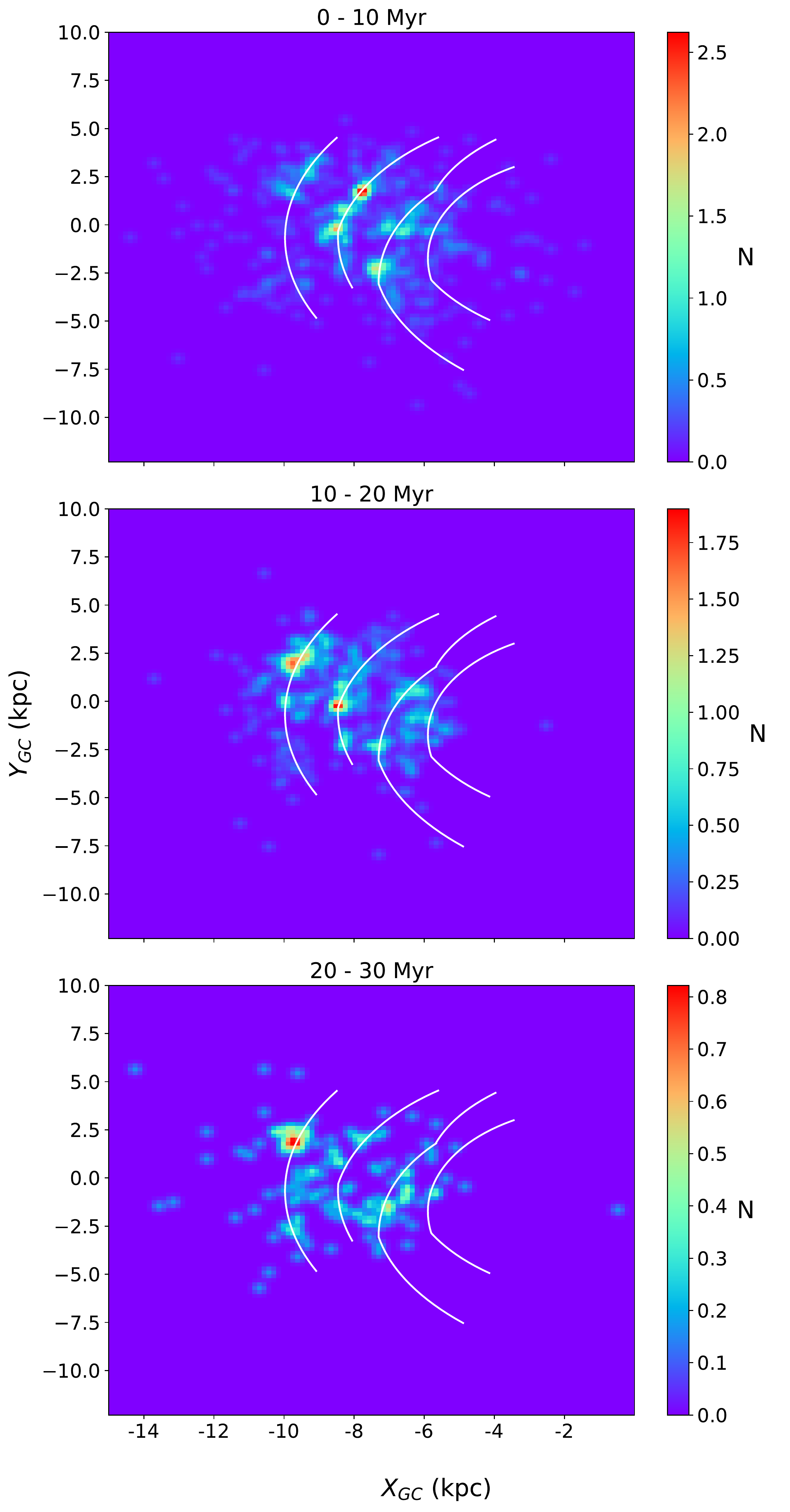}
\caption{Density heat map for OCs in different age intervals of \(\leq\) 10 Myr, 10 - 20 Myr, and 20 - 30 Myr (from top to bottom). Spiral arms are traced by the OCs younger than 10 Myr and the older ones occupy the inter-arm regions. Further, the spiral structure loses its shape in the 20 - 30 Myr age range.}
\label{fig03}
\end{figure}
%
%
\subsection{Delineation of Spiral arms}
The demarcation of clusters in terms of age band is crucial for untangling the spiral arm structure, as it has been known for a long time that young OCs are ideal tracers of the Galactic spiral arms. In Figure~\ref{fig03}, we plot the distribution for three different samples of OCs based on their ages viz., (i) \(Age/\text{Myr} \leq 10 \hspace{0.5mm}\) (top panel), (ii) \(10 \hspace{0.5mm} < Age/\text{Myr} \leq 20 \hspace{0.5mm}\) (middle panel), and (iii) \(20 \hspace{0.5mm} < Age/\text{Myr} \leq 30 \hspace{0.5mm}\) (bottom panel). This displays the respective locations of 501, 355, and 141 OCs in the $X-Y$ plane. The plots depict spiral arms defined by \citet{2021FrASS...8..103H} overplotted as curved solid lines in each of the subplots of Figure~\ref{fig03}. The importance of one or more kinks, as also proposed by \citet{2019ApJ...885..131R} and \citet{2021FrASS...8..103H}, in the spiral arms becomes apparent in the top panel of the Figure~\ref{fig03}, especially in the case of the Local and Sagittarius-Carina Arms. The spiral arms deduced by \citet{2021FrASS...8..103H} have allowed us an extended view of the Milky Way spiral arms for a more extensive range of galactic azimuth in the negative direction in comparison to the earlier studies of \citet{2014ApJ...783..130R} and \citet{2021A&A...652A.162C}. One can observe that the younger OCs (Age \(<\) 10 Myr) clearly trace out the spiral arms in the solar neighbourhood and occupy the inter spiral arm regions as they age, while most of the OCs having their age in the range 20-30 Myr are well off from the spiral arms. This means that, in general, most of the OCs leave their birth site after about 10 to 20 Myr and start populating in the inter-arm regions where they stay for the rest of their life. \citet{2021A&A...652A.162C} has also noted, however, with different age bins, that the OC over-densities showed increased dispersion with time. Moreover, it was also observed that many clusters do not survive longer than about 10-15 Myr after their birth with the death rate decreasing progressively as they become older and leave the spiral arms \citep{2021FrASS...8...62M}. 
%
\begin{figure}
\centering
\includegraphics[width=8.5cm,height=7.5cm]{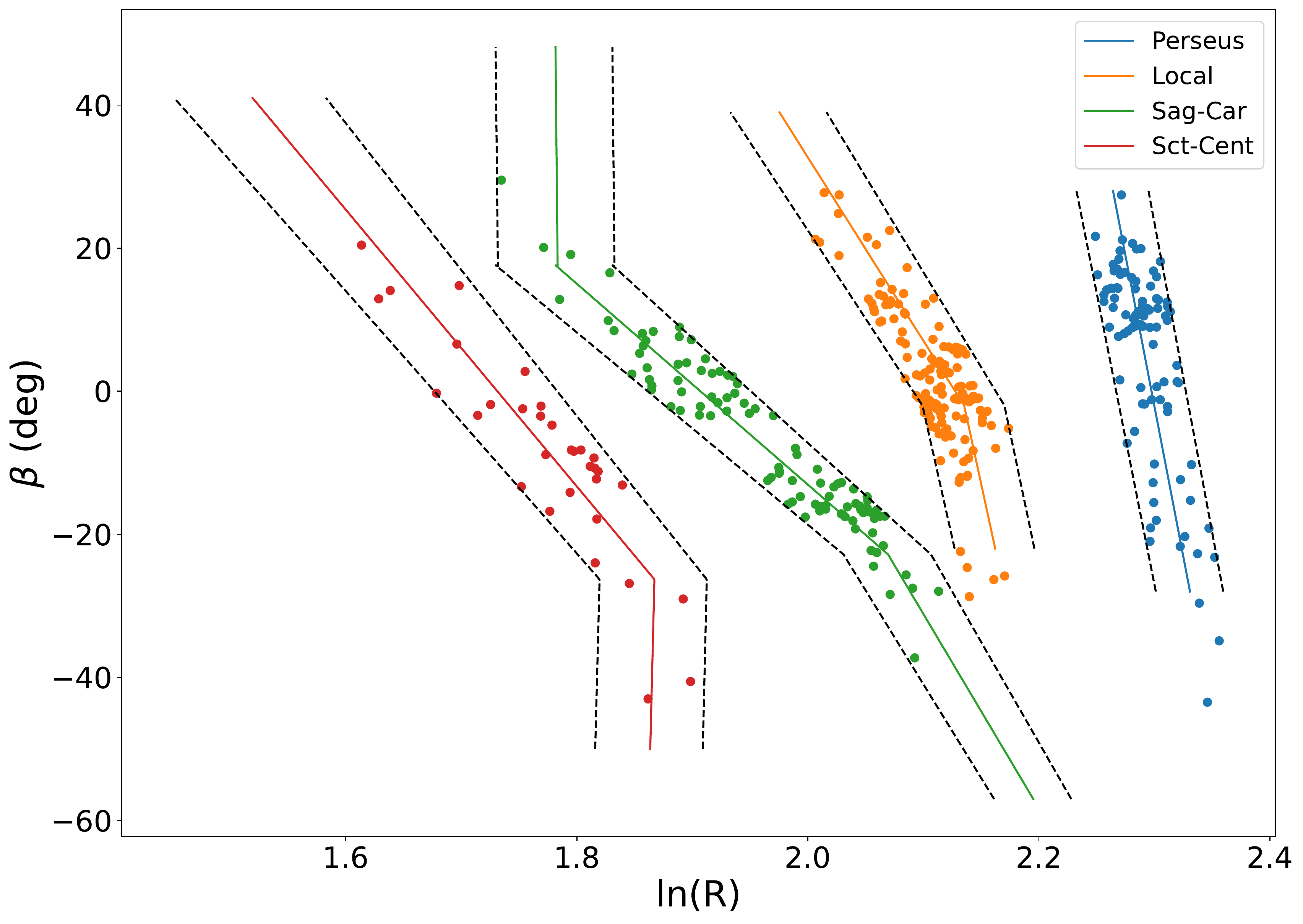}
\caption{Distribution of OCs in ln(R) - \(\beta\) plane after arm assignments. The solid lines denote the spiral arms taken from \citet{2021FrASS...8..103H}, whereas the dashed lines indicate a width of 0.3 kpc for each spiral arm.}
\label{fig04}
\end{figure}
\subsection{Spiral Pattern rotation speed}
Determining spiral pattern rotation speeds can provide us with critical insights into the nature and evolution of spiral arms. Since the parameters of OCs can be determined with high precision by averaging over the members, one can make use of their kinematic data to estimate the pattern speed of spiral arms \citep{2005ApJ...629..825D, 2011MSAIS..18..185G}. Some recent studies on the determination of spiral pattern rotation speeds include \citet{2021A&A...652A.162C}, which have made use of 264 young OCs, and \citet{2022MNRAS.512.1169G} which analysed 252 OCs within three kpc of the Sun and having age \(<\) 100 Myr. Taking advantage of a larger sample of OCs available in the present study, we re-investigated the rotation speeds of spiral arms in the following analysis.

\begin{figure}
\centering
\includegraphics[width=8.5cm,height=7.5cm]{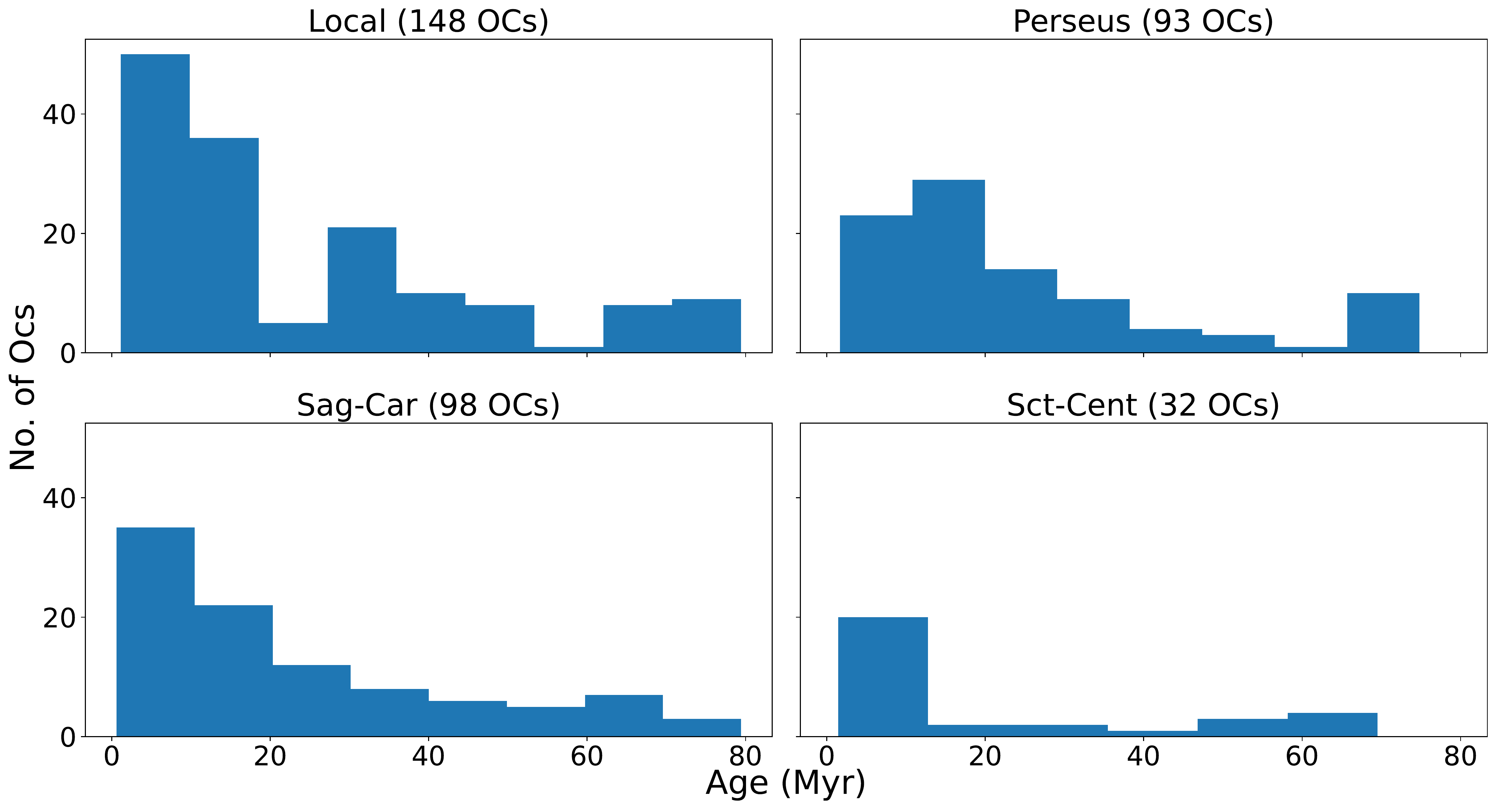}
\caption{Age distribution of OCs belonging to each spiral arm. The number of clusters assigned to each arm is mentioned in parentheses. As expected the counts of observed clusters decrease as the age increases.}
\label{fig05}
\end{figure}

Before proceeding further, we needed to identify the host arms for the OCs. This was done by the arm assignments of the OCs lying within 0.3 kpc on either side of their nearest arm. Figure~\ref{fig04} shows each arm and the corresponding OCs in the ln(R) - \(\beta\) plane, where $R$ and \(\beta\) are the same as defined in the previous section. The advantage of showing OCs in ln(R) - \(\beta\) plane is that the spiral arms in this plot can be described with a set of one or more straight lines having different slopes at the positions of kinks which is helpful in separating the OCs that are currently part of a particular spiral arm. The arm assignment was unambiguous in most cases and was supplemented with the kinematic data of OCs in the case of clusters near the borders. We also included the OCs, which were not in the \(\beta\) range of present-day spiral arms but were part of their appropriate extensions, as seen in the case of the Local and the Perseus Arms in the figure. We have taken the solar distance from the Galactic center \(R_\odot\) as 8.15 kpc and the circular rotation speed at the solar position \(\Theta_{0} = 236 \pm 7\) km \(s^{-1}\) \citep{2019ApJ...885..131R}. We restricted the OCs age within \(\leq\) 80 Myr, yielding us a total sample of 371 OCs having 93, 148, 98, and 32 OCs in Perseus, Local, Sagittarius-Carina (Sag-Car), and Scutum-Centaurus (Sct-Cent) arms, respectively which also have the information available for their radial velocities. As mentioned in Section \ref{data}, due to very few OCs with estimated distance uncertainties, we only use the estimated value of the distance for arm assignments of OCs. Therefore, the results of the arm assignments are subject to change in future OC catalogs with slightly different set of values for the physical parameters of OCs. This reflects on the limitation of the inhomogeneous nature of our catalog. However, a relatively small mean uncertainty of 170 pc in distance reported by \citet{2021MNRAS.504..356D} using the stellar membership of 1743 OCs determined from Gaia DR2 astrometric data leads us to the conclusion that our subsequent results present an unblemished view of the Milky Way. Figure~\ref{fig05} shows the age distribution of OCs in each arm, where we observe a gradual decline in the number of OCs as the age increases. This might be caused due to three factors.
\begin{enumerate}
\item With the evolution of clusters in the molecular gas clouds, only about 5\% survive by the time they get detached from the parent molecular cloud  \citep{2003ARA&A..41...57L}.
\item Older OCs might have undergone disruption due to close encounters with other clusters and clouds of gases. 
\item OCs might have moved out of the spiral arms via relative displacement in both the Galactic plane and in a vertical direction away from the Galactic mid-plane in such a way that they are no longer considered to be a part of the spiral arms \citep{2011A&A...527A.147M, 2020MNRAS.495.2673C, 2021ApJ...919...52Z}.
\end{enumerate}
%

Since OCs younger than 10 Myr trace the Milky Way spiral arms quite well and only begin to move away after this time, we used the OCs older than 10 Myr to estimate the rotational speed of the spiral arms. We adopted a similar methodology proposed by \citet{2005ApJ...629..825D} to compute the pattern speed \(\Omega\) corresponding to each arm:
\begin{itemize}
\item Arm assignment was done for each OC and each arm was studied separately.
\item Birthplaces were found by integrating the orbit of each OC backward and it was assumed that these birthplaces represent points on the spiral arm time $T$ ago where $T$ denotes the age of each cluster.
\item Once we had the birthplaces of OCs, we assumed that the spiral arms traversed a circular path in the Galactic $X-Y$ plane with time and calculated the present-day Galactic azimuths for a particular pattern speed \(\Omega\) given the Galactic azimuths corresponding to the birthplaces of OCs. Mathematically,
\[\beta_{G, present} = \beta_{G, birth} + \Omega \cdot T\]
\item We estimated the optimal value of \(\Omega\) by minimising the distance between the present-day spiral arms and their integrated present-day locations. The mean of the pattern speeds obtained for OCs in a spiral arm was reported as its best value \(\Omega\) with its dispersion represented by the standard deviation.
\end{itemize}
We performed the OC orbit integration following the MWPotential2014 of the Python package GALPY \citep{2015ApJS..216...29B} which is fit to various observational constraints and is composed of a spherical nucleus and a bulge, Navarro-Frenk-White dark matter halo and a Miyamoto-Nagai disc. The numerical integration was done using the Leapfrog integration scheme where the orbits were traced back in steps of 0.1 Myr. Also, we have rejected the data points on the integrated spiral arms that did not lie within 0.3 kpc of the analytical spiral arms, wherein the number of rejected data points varies on using different values for \(\Omega\). Since uncertainty in age is the dominant source of error in the computed birthplaces, we estimated the typical value of the error in age, whenever available, to be about 20\(\%\) which is close to the value of 22\(\%\) average age error for a sample of 581 OCs reported by \citet{2004NewA....9..475D}. Hence, we iterated the above process 100 times by using the age values of clusters with corresponding random errors ranging from 0 - 20\(\%\). The mean value and the standard deviation are reported in Table~\ref{tab:2} as the optimal and the corresponding dispersion in the pattern speed \(\Omega\). 

\begin{table*}
  \centering
  \caption{Spiral Arm rotation speeds for four different spiral arms and three different age intervals}
  \begin{tabular}{lcccr}
  \hline
    Arm & No. of & \multicolumn{3}{c}{$\Omega_{rot}$ [km s\(^{-1}\) kpc\(^{-1}\)]} \\
    &  OCs & (10 - 80 Myr) & (10 - 50 Myr) & (50 - 80 Myr)\\
    \hline 
    Perseus & 93 & 23.09 \(\pm\) 0.82 & 18.97 \(\pm\) 0.76 & 25.78 \(\pm\) 1.13\\
    Local   & 148 &  31.47 \(\pm\) 0.55 & 32.86 \(\pm\) 0.64  & 30.32 \(\pm\) 1.12\\
    Sag-Car & 98 & 30.9 \(\pm\) 0.59 & 29.56 \(\pm\) 0.94  & 31.97 \(\pm\) 1.2 \\
   Sct-Cent &  32 & 52.99 \(\pm\) 2.34  & 51.99 \(\pm\) 4.15  & 53.02 \(\pm\) 3.54 \\
    \hline
  \end{tabular}
  \label{tab:2}
\end{table*}

We computed the pattern speeds of each spiral arm for different age bins to investigate if the spiral arms have accelerated in the last 80 Myrs. The resultant values are given in Table \ref{tab:2} where the rotation speed of each arm is provided in the units of km s\(^{-1}\) kpc\(^{-1}\) along with the number of OCs younger than 80 Myr assigned to each spiral arm. Here, one has to note that the estimated errors in the pattern speeds represent only a lower limit since we could not account for the errors in 3D kinematics and distance measurements due to their unavailability in the reference catalogs. Moreover, \cite{2021A&A...652A.162C} performed a simulation that aimed at examining if the proposed methodology can be used to distinguish two different values of \(\Omega_{rot}\) (in their case 20 and 50 km s\(^{-1}\) kpc\(^{-1}\)). As a result, they obtained a typical systematic error of 0.8 to 6 km s\(^{-1}\) kpc\(^{-1}\) in the imposed pattern speed and were able to conclude that though the methodology for the determination of spiral arm pattern speed is not accurate enough to estimate the exact pattern speeds of individual spiral arms, it is sufficiently good to be able to differentiate between two completely distant values of pattern speeds. Keeping the above facts in mind, one can deduce the following arguments from the estimated pattern speeds given in Table \ref{tab:2}:

\begin{enumerate}
    \item On comparing the pattern speeds of a particular spiral arm across different age bins, it is evident that the spiral arms have not accelerated in the last 80 Myrs.
    \item In a particular age bin, the estimated pattern speeds of the four spiral arms follow a declining trend as a function of the galactocentric radius as seen in Fig. \ref{fig05}, except for the case of the Local Arm, which has a slightly higher mean pattern speed than Sag-Car Arm.
    \item Though there have been recent indications that the Local Arm is larger than previously thought with its pitch angle comparable to Galaxy's other major arms (\cite{2016SciA....2E0878X}), this contrasting behaviour requires further investigation. A similar observation was also made by \cite{2021A&A...652A.162C}, though we cannot compare the exact values of the pattern speeds in two studies due to different values of \(R_\odot\) and \(V_\odot\) used to normalize the rotation curve.
\end{enumerate}
%
\begin{figure}
\vspace{0.5cm}
\centering
\includegraphics[width=8.5cm,height=7.0cm]{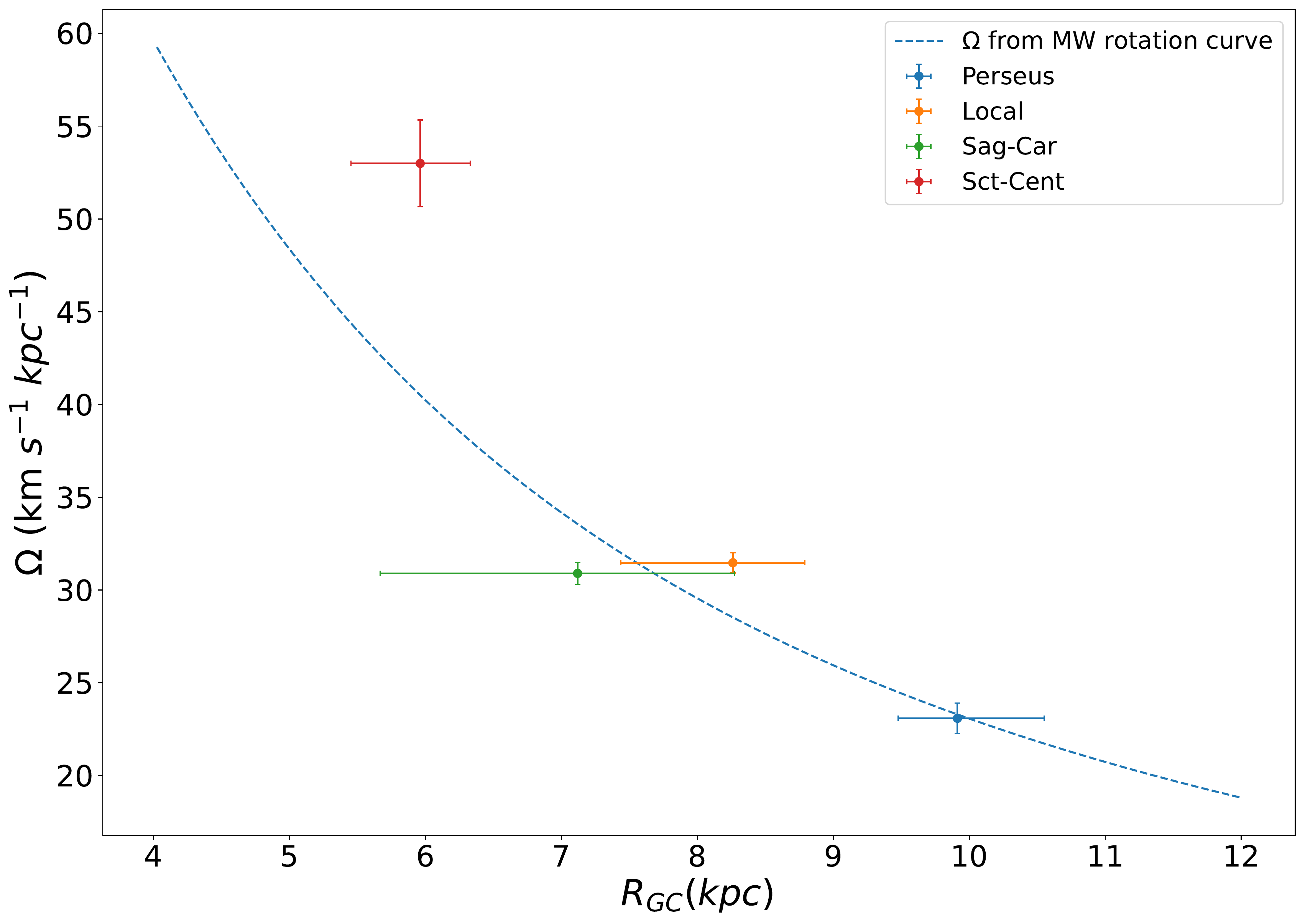}
\caption{Spiral Arm pattern speeds corresponding to the 10 - 80 Myr interval estimated for the Sct-Cent, Sag-Car, Local, and the Perseus Arm shown along with the angular velocity obtained from the MWPotential2014 rotation curve as a function of the galactocentric distance. The \(R_{GC}\) values for the solid data points and the associated horizontal error bars denote the mean galactocentric distance and its spread for the distribution of OCs in each spiral arm, respectively.}
\label{fig06}
\end{figure}
%
Our results are in favour of a flocculent Milky Way with short-lived transient spiral arms wherein the spiral arms co-rotate with the Milky Way disk at every point \citep{2012MNRAS.421.1529G, 2014MNRAS.443.2757K}. This is in substantial agreement with the recent works done by \citet{2018MNRAS.481.3794H, 2018MNRAS.480.3132Q} who have been able to explain particular ridge and arc-like features in the phase space distributions of stars in the local neighborhood by winding spiral arm models with different pattern speeds at different galactocentric radii. However, our results are in stark contrast with the work done by \cite{2019MNRAS.486.5726D} and \cite{2021FrASS...8...62M}, who adopt the same methodology and obtain the same pattern speed for all spiral arms while supporting the density wave scenario of the Milky Way spiral arms. Further, several numerical N-body simulations (\cite{2006MNRAS.367..873D}; \cite{2011MNRAS.410.1637S}; \cite{2011ApJ...735....1W}; \cite{2012ApJ...745L..14P}; \cite{2013MNRAS.432.2878R}; \cite{2018MNRAS.481..185M}) lead to a transient behaviour of spiral arms corotating with the disk stars in case of unbarred or weak bar galaxies. Moreover, a lack of an age gradient of clusters across the spiral arms (\cite{2010MNRAS.409..396D}; \cite{2021A&A...652A.162C}) also points in a similar direction.

\vspace{0.5cm}
\section{Distribution of OCs perpendicular to the Galactic plane}\label{cluster_distri}
\begin{figure}
\centering
\includegraphics[width=8.5cm, height=7.0cm]{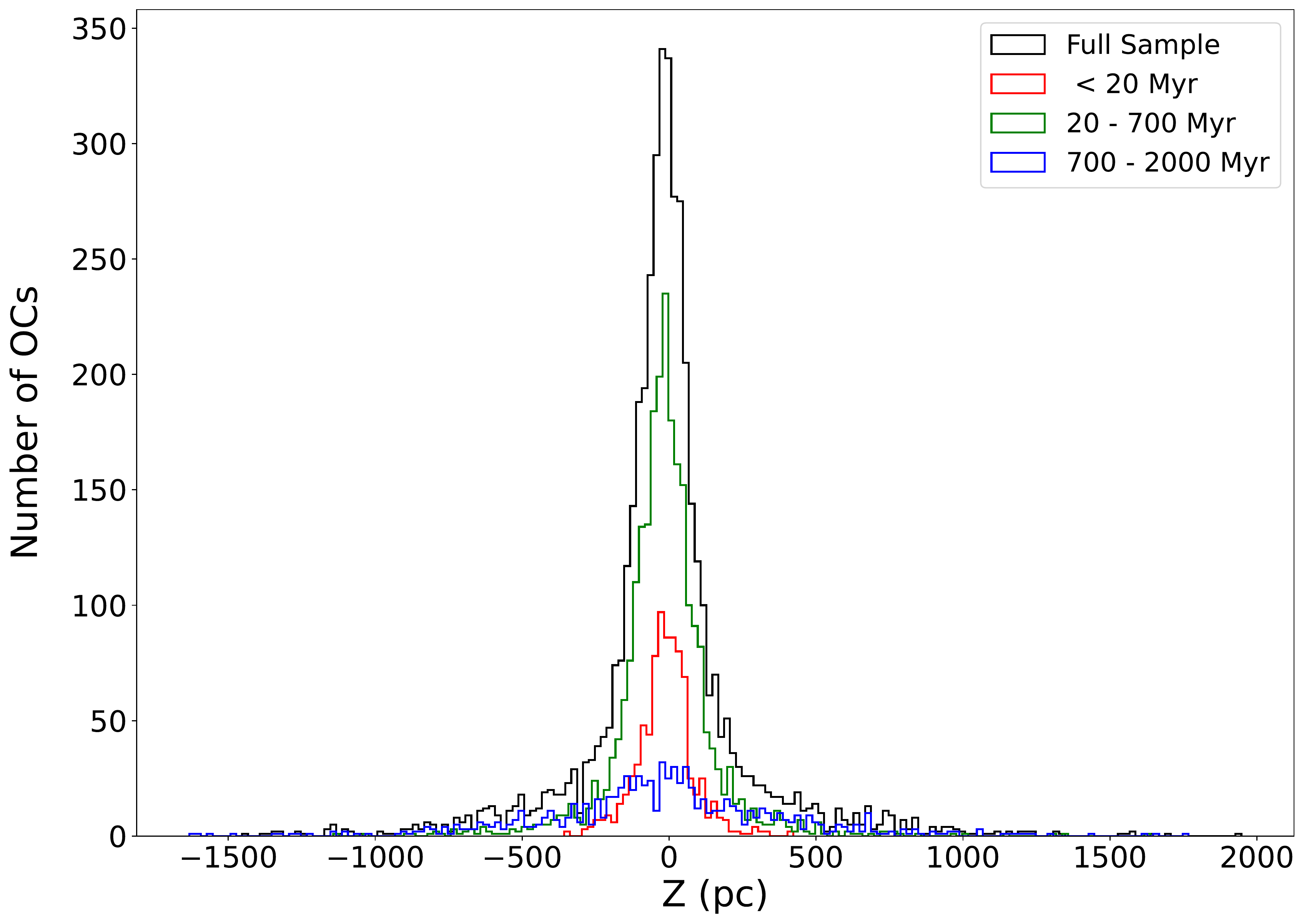}
\caption{Z distribution of OCs corresponding to different age intervals. The scale height is estimated while assuming a constant solar offset \(z_\odot\) of \(17.0 \pm 0.9\) pc for all the plots. The sample is restricted to (-2, 2) kpc in $z$. The colors corresponding to each OC sample are mentioned in the plot's top right corner.}
\label{fig07}
\end{figure}

For a symmetric Galaxy, the number of OCs below and above the Galactic plane is expected to be the same, and the number density of the OCs is maximum near the Galactic plane that falls steadily as one goes away from the mid-plane. However, it is known that there is an asymmetry in the OCs distribution around the Galactic mid-plane. In Figure~\ref{fig07}, which shows the vertical distance distribution (\(z = d\,sin\,b\)), where d is the heliocentric distance and b is the Galactic latitude), we divided the cluster sample into different bins in $z$ with a bin-width of 20 pc, wherein the number distribution peaks at \( z \sim -15\) pc. Since there were very few OCs at more considerable vertical distances, we have restricted $z$ for OCs within two kpc from the Galactic plane, up to which the OCs sample is generally complete. This has left us with 4372 OCs used in the following analysis.
\subsection{Solar offset}\label{offset}
It is commonly assumed that the cluster number density distribution perpendicular to the Galactic plane can be well described in the form of a decaying exponential away from the Galactic plane, as given by \citet{1998A&A...336..137C},
$$
N = N_0 exp\left[-\frac{|z+z_\odot|}{z_h}\right] 
$$
where $z_\odot$ and $z_h$ are the solar offset and scale height, respectively. Here, \(N_0\) is the maximum cluster density, which by definition, is attained at a vertical height \(z = - z_\odot\). The knowledge of the exact value of $z_\odot$ is essential not just for the Galactic structure models but also in describing the north-south asymmetry in the density distribution of different kinds of stellar populations in the north and south Galactic regions \citep{1995ApJ...444..874C, 1999A&A...352..459C, 2005MNRAS.362.1259J, 2016AstL...42....1B}. However, to accurately determine the solar offset, we must consider the OCs close to the Galactic plane to prevent any bias induced by far-away clusters. To do this, we applied the age cut-off for OCs as the clusters start drifting away from the Galactic mid-plane as they age. We refer the reader to Figure~2 of \citet{2007MNRAS.378..768J}, and on using the same methodology, we only considered the OCs younger than \(\sim\) 700 Myr. We determined $z_\odot$ by fitting the above exponential function to the sample of OCs younger than 700 Myr and obtained an estimate of \(z_\odot = 17.0 \pm 0.9\) pc. 

Numerous studies have estimated $z_\odot$ using a wide variety of celestial sources, and most of these studies obtained $z_\odot$ in the range of 10 to 25 pc in the north direction of Galactic plane \citep[e.g.,][]{2007MNRAS.378..768J, 2009MNRAS.398..263M, 2014MNRAS.444..290B, 2017MNRAS.468.3289Y, 2019A&A...632L...1S, 2020AA...640A...1C, 2021MNRAS.502.4194G}. Though a direct comparison of $z_\odot$ in different studies is difficult, it is found that the choice of the data sample and method of determination account for most of the disagreements among $z_\odot$ values \citep{2007MNRAS.378..768J, 2020AA...640A...1C}. \citet{2019ApJ...885..131R} attributed these differences  as a combination of extinction and Galactic warping. It is observed that $z_\odot$ derived through young tracers like H II regions, masers, and molecular clouds are generally lower than that found using the stellar tracers like Classical Cepheids, RR Lyrae stars, and star clusters \citep[e.g.,][]{2016AstL...42....1B, 2016AstL...42..182B, 2017MNRAS.465..472K}. Nevertheless, our present estimate of $z_\odot$ is in remarkable agreement with \citet{2016AstL...42....1B} mean value of \(16\pm2\) pc and \citet{2017MNRAS.465..472K} median value of \(17\pm2\) pc. 
%
%
\subsection{Cluster scale height}
The observed clusters distribution perpendicular to the Galactic plane provides us with an estimate of the thickness of the young Galactic disk. We fitted the number distribution of OCs in the exponential profile as described in Eq. (1) to estimate the scale height wherein the best fit for the sample of 4372 OCs resulted in a scale height of $z_h$ = 100.8 \(\pm\) 1.7 pc. However, the large value of $z_h$ is mainly influenced by the older clusters, mainly located at higher latitudes resulting in a shallower wing in the $z$ distribution. We plot the similar distribution of OCs perpendicular to the Galactic plane but using different age bins viz., \(<\) 20 Myr (red), 20 - 700 Myr (green), 700 - 2000 Myr (blue) corresponding to 837, 2439, and 831 OCs respectively. We fitted the exponential profile in all these three samples to estimate the scale height while keeping the solar offset $z_\odot = 17.0$ pc as a fixed value as derived in the previous section. In addition, we also determined scale heights for the larger cluster sample having age \(<\) 700 Myr and \(<\) 2000 Myr. Table~\ref{tab:3} provides information about the scale heights obtained for different age bins and the corresponding number of OCs. The scale height increases as we fit the exponential distribution to older OCs, as also evident in Fig. \ref{fig07}, which is consistent with the hypothesis that OCs migrate toward the outer disk as they age. The pattern speed of spiral arms depicts a decreasing trend with increasing galactocentric distance. For a sample of OCs younger than 700 Myr, $z_h$ is obtained as $91.7 \pm 1.9$ pc. A range of cluster scale heights has been published by the authors in previous studies. For example, \citet{2006A&A...446..121B} derived a scale height of 48 \(\pm\) 3 pc for OCs younger than 200 Myr; \citet{2016A&A...593A.116J} obtained \(z_h = 64 \pm 2\) pc for OCs within 1.8 kpc of the Sun. \citet{2014MNRAS.444..290B} found the scale height increases from about 40 pc at 1 Myr to 75 pc at the age of 1 Gyr. Some more recent studies, such as \citet{2020AA...640A...1C} obtained a scale height of 74 \(\pm\) 5 pc for OCs in the solar neighbourhood with a typical age of \(\sim\) 100 Myr. Similarly, \citet{2021AA...652A.102H} reported a scale height of 70.5 \(\pm\) 2.3 pc for young OCs (Age \(<\) 20 Myr), which increases to 87.4 \(\pm\) 3.6 pc for older OCs with ages of 20 - 100 Myr. Although it is difficult to compare different estimates due to their different sample selection criteria, our present value of $z_h = 91.7 \pm 1.9$ pc for clusters younger than \(<\) 700 Myr agrees with most of these recent estimates. 
\begin{table}
  \centering
  \caption{Estimated scale heights for different age intervals of the cluster sample while keeping the solar offset \(Z_\odot = 17.0 \pm 0.9\) pc fixed.}
  \begin{tabular}{lcr}
  \hline
    Age Interval (Myr) & \(Z_{h}\) (pc)  & No. of OCs \\
    \hline 
     Full Sample & 100.8 \(\pm\) 1.7 & 4372 \\
    \(<\) 20   & 80.8 \(\pm\) 8.0 & 837 \\
    \(<\) 700  & 91.7 \(\pm\) 1.9 & 3376 \\
    \(<\) 2000 & 98.8 \(\pm\) 1.6 & 4107 \\
    20 - 700   & 96.4 \(\pm\) 1.9 & 2439\\
    700 - 2000 &294.7 \(\pm\) 19.5& 831\\
    \hline
  \end{tabular}
  \label{tab:3}
\end{table}

Further, it has also been established that the OCs farther away from the Galactic center are present at higher altitudes from the Galactic plane hence resulting in the variation of $z$ scale height as a function of the projected Galactocentric distance (\(R_{GC}\)). \citet{2006A&A...446..121B} observed that the scale height increases by a factor of two when one compares the two estimates for OCs inside and outside the solar circle. \citet{2014MNRAS.444..290B} and \citet{2016A&A...593A.116J} showed a positive trend between the galactocentric distance and the scale height. Using our OC sample, we calculated the scale height (\(z_{h}\)) for two cases (i) \(R_{GC}\) \(\leq\) 8.15 kpc (ii) \(R_{GC}\) \(>\) 8.15 kpc comprised of 1865 and 2513 OCs respectively. Here, 8.15 kpc is taken as the location of solar orbit \citep{2019ApJ...885..131R}. We obtained \(z_{h}\) to be 79.2 \(\pm\) 1.6 pc and 138.1 \(\pm\) 1.7 pc for the two cases, respectively, which further affirms dependence of \(z_{h}\) on the Galactocentric distance, in agreement with both these studies.
\section{Summary}\label{summary}
In this work, we compiled various catalogs of Galactic OCs that became available recently and found a total of 6133 OCs, most of which have been detected in the post-Gaia phase using the photometric and kinematic information provided in the Gaia archive data. Since all open clusters, including the progenitors of giant molecular clouds, are believed to be initiated their lives in the Galactic disk, they have been widely used in the study of the structure of the Galactic disk. We used the position, distance, age, and kinematic parameters of these clusters in the Milky Way to study the Galactic disk morphology. For our work, we rely on the catalogs provided in more than 35 references published in the past. The ages of OCs cover a whole range, from a few Myrs to tens of billions of years, making them excellent tracers for investigating the Galactic disk, spiral structure, and in particular the evolution of spiral arms. Though our cluster sample seems complete only up to 2 kpc from the Sun, clusters are found to be located as far as 13 kpc from the Solar position. 

As the young open clusters remain close to their birthplace and, in most cases, are associated with star-forming regions, these are ideal tracers to probe the spiral arms of the Galaxy. The distribution of our sample of OCs in the 2-D plane clearly showed signatures of the spiral arms, mainly when we restricted the cluster sample to less than 30 Myrs, as older OCs have a more scattered distribution. In the spiral arms and inter-arm regions, the distributions of OCs reveal the complex substructures in the spiral arms. By analysing young clusters in different age bins, we could delineate the spiral structures and found that clusters start leaving the spiral arms after 10-20 Myrs and fill the inter-arm regions where they stay for the rest of their life. The distribution of these young clusters traced a four-arm spiral in the Galaxy. The two spiral structures related to the Local arm and Sagittarius-Carina spiral arms were conspicuous by their presence. In contrast, some substructures of Perseus and Scutum-Centaurus spiral arms were feebly present. Once we could trace the spiral arms using the younger clusters, we studied their pattern speed by adopting the methodology given by \citet{2005ApJ...629..825D}. With the help of a large sample of OCs with 3D kinematics, we found that the spiral arms rotate with different individual pattern speeds, which tend to follow a declining trend as one moves towards an increasing galactocentric radius. This suggests that the Milky Way spiral arms might be transient in nature, thereby favouring a flocculent Milky Way. On comparing the pattern rotation speeds of spiral arms across different age bins, it is found that the spiral arms rotation speeds have not changed much in the last 80 Myr.

The number density distribution of OCs has been used to determine the Solar offset and scale height, which are about $17.0\pm0.9$ pc and $91.7 \pm 1.9$ pc, respectively, for the clusters younger than 700 Myrs. However, we found these numbers vary slightly when we alter the selection criteria of the chosen cluster sample in terms of age, distance perpendicular to the Galactic plane, and Galactocentric distance of the clusters sample. In the near future, more improved parallax and distance estimates are expected to be available in the Gaia DR4 survey, providing a larger sample of open clusters having precise physical parameters. This in turn will allow better in-depth study of spiral arm features and constraints on the formation of spiral arms in the Galaxy.
\section*{Acknowledgments}
We thank referee for the suggestions which has improved the quality of this paper. SM is grateful to ARIES for providing support through its Visiting Student Program (VSP) to conduct his master's thesis project. We would also like to thank Deepak for his critical remarks.  This work has made use of data from the European Space Agency (ESA) mission Gaia (https://www.cosmos.esa.int/gaia), processed by the Gaia Data Processing and Analysis Consortium (DPAC, https://www.cosmos.esa.int/web/gaia/dpac/consortium). 

\bibliography{main}{}

\begin{thebibliography}{}
\expandafter\ifx\csname natexlab\endcsname\relax\def\natexlab#1{#1}\fi
\providecommand{\url}[1]{\href{#1}{#1}}

\bibitem[{{Bobylev} \& {Bajkova}(2016{\natexlab{a}})}]{2016AstL...42....1B}
{Bobylev}, V.~V., \& {Bajkova}, A.~T. 2016{\natexlab{a}}, Astronomy Letters,
  42, 1

\bibitem[{{Bobylev} \& {Bajkova}(2016{\natexlab{b}})}]{2016AstL...42..182B}
---. 2016{\natexlab{b}}, Astronomy Letters, 42, 182

\bibitem[{{Bonatto} {et~al.}(2006){Bonatto}, {Kerber}, {Bica}, \&
  {Santiago}}]{2006A&A...446..121B}
{Bonatto}, C., {Kerber}, L.~O., {Bica}, E., \& {Santiago}, B.~X. 2006, \aap,
  446, 121

\bibitem[{{Bossini} {et~al.}(2019){Bossini}, {Vallenari}, {Bragaglia},
  {Cantat-Gaudin}, {Sordo}, {Balaguer-N{\'u}{\~n}ez}, {Jordi}, \&
  et~al.}]{2019AA...623A.108B}
{Bossini}, D., {Vallenari}, A., {Bragaglia}, A., {et~al.} 2019, \aap, 623, A108

\bibitem[{{Bovy}(2015)}]{2015ApJS..216...29B}
{Bovy}, J. 2015, \apjs, 216, 29

\bibitem[{{Buckner} \& {Froebrich}(2013)}]{2013MNRAS.436.1465B}
{Buckner}, A. S.~M., \& {Froebrich}, D. 2013, \mnras, 436, 1465

\bibitem[{{Buckner} \& {Froebrich}(2014)}]{2014MNRAS.444..290B}
---. 2014, \mnras, 444, 290

\bibitem[{Cantat-Gaudin(2022)}]{universe8020111}
Cantat-Gaudin, T. 2022, Universe, 8

\bibitem[{{Cantat-Gaudin} \& {Anders}(2020)}]{2020AA...633A..99C}
{Cantat-Gaudin}, T., \& {Anders}, F. 2020, \aap, 633, A99

\bibitem[{{Cantat-Gaudin} {et~al.}(2020){Cantat-Gaudin}, {Anders},
  {Castro-Ginard}, {Jordi}, {Romero-G{\'o}mez}, \& et~al.}]{2020AA...640A...1C}
{Cantat-Gaudin}, T., {Anders}, F., {Castro-Ginard}, A., {et~al.} 2020, \aap,
  640, A1

\bibitem[{{Cantat-Gaudin} {et~al.}(2018){Cantat-Gaudin}, {Jordi}, {Vallenari},
  {Bragaglia}, \& et~al.}]{2018AA...618A..93C}
{Cantat-Gaudin}, T., {Jordi}, C., {Vallenari}, A., {Bragaglia}, A., \& et~al.
  2018, \aap, 618, A93

\bibitem[{{Cantat-Gaudin} {et~al.}(2019){Cantat-Gaudin}, {Krone-Martins}, \&
  et~al.}]{2019A&A...624A.126C}
{Cantat-Gaudin}, T., {Krone-Martins}, A., \& et~al. 2019, \aap, 624, A126

\bibitem[{{Carrera} {et~al.}(2019){Carrera}, {Bragaglia}, {Cantat-Gaudin}, \&
  et~al.}]{2019AA...623A..80C}
{Carrera}, R., {Bragaglia}, A., {Cantat-Gaudin}, T., \& et~al. 2019, \aap, 623,
  A80

\bibitem[{{Casamiquela} {et~al.}(2021){Casamiquela}, {Soubiran}, {Jofr{\'e}},
  {Chiappini}, {Lagarde}, \& et~al.}]{2021AA...652A..25C}
{Casamiquela}, L., {Soubiran}, C., {Jofr{\'e}}, P., {et~al.} 2021, \aap, 652,
  A25

\bibitem[{{Castro-Ginard} {et~al.}(2019){Castro-Ginard}, {Jordi}, {Luri},
  {Cantat-Gaudin}, \& {Balaguer-N{\'u}{\~n}ez}}]{2019A&A...627A..35C}
{Castro-Ginard}, A., {Jordi}, C., {Luri}, X., {Cantat-Gaudin}, T., \&
  {Balaguer-N{\'u}{\~n}ez}, L. 2019, \aap, 627, A35

\bibitem[{{Castro-Ginard} {et~al.}(2022{\natexlab{a}}){Castro-Ginard}, {Jordi},
  {Luri}, {Cantat-Gaudin}, \& et~al.}]{2022AA...661A.118C}
{Castro-Ginard}, A., {Jordi}, C., {Luri}, X., {Cantat-Gaudin}, T., \& et~al.
  2022{\natexlab{a}}, \aap, 661, A118

\bibitem[{{Castro-Ginard} {et~al.}(2021){Castro-Ginard}, {McMillan}, {Luri},
  {Jordi}, \& et~al.}]{2021A&A...652A.162C}
{Castro-Ginard}, A., {McMillan}, P.~J., {Luri}, X., {Jordi}, C., \& et~al.
  2021, \aap, 652, A162

\bibitem[{{Castro-Ginard} {et~al.}(2022{\natexlab{b}}){Castro-Ginard}, {Jordi},
  {Luri}, {Cantat-Gaudin}, {Carrasco}, {Casamiquela}, {Anders},
  {Balaguer-N{\'u}{\~n}ez}, \& {Badia}}]{2022A&A...661A.118C}
{Castro-Ginard}, A., {Jordi}, C., {Luri}, X., {et~al.} 2022{\natexlab{b}},
  \aap, 661, A118

\bibitem[{{Chen} {et~al.}(1999){Chen}, {Figueras}, {Torra}, {Jordi}, {Luri}, \&
  {Galad{\'\i}-Enr{\'\i}quez}}]{1999A&A...352..459C}
{Chen}, B., {Figueras}, F., {Torra}, J., {et~al.} 1999, \aap, 352, 459

\bibitem[{{Chen} {et~al.}(1998){Chen}, {Vergely}, {Valette}, \&
  {Carraro}}]{1998A&A...336..137C}
{Chen}, B., {Vergely}, J.~L., {Valette}, B., \& {Carraro}, G. 1998, \aap, 336,
  137

\bibitem[{{Chen} {et~al.}(2003){Chen}, {Hou}, \& {Wang}}]{2003AJ....125.1397C}
{Chen}, L., {Hou}, J.~L., \& {Wang}, J.~J. 2003, \aj, 125, 1397

\bibitem[{{Chen} \& {Zhao}(2020)}]{2020MNRAS.495.2673C}
{Chen}, Y.~Q., \& {Zhao}, G. 2020, \mnras, 495, 2673

\bibitem[{{Chi} {et~al.}(2023){Chi}, {Wang}, {Wang}, {Deng}, \&
  {Li}}]{2023ApJS..266...36C}
{Chi}, H., {Wang}, F., {Wang}, W., {Deng}, H., \& {Li}, Z. 2023, \apjs, 266, 36

\bibitem[{{Cohen}(1995)}]{1995ApJ...444..874C}
{Cohen}, M. 1995, \apj, 444, 874

\bibitem[{{de la Fuente Marcos} \& {de la Fuente
  Marcos}(2004)}]{2004NewA....9..475D}
{de la Fuente Marcos}, R., \& {de la Fuente Marcos}, C. 2004, \na, 9, 475

\bibitem[{{Dias} {et~al.}(2002){Dias}, {Alessi}, {Moitinho}, \&
  {L{\'e}pine}}]{2002AA...389..871D}
{Dias}, W.~S., {Alessi}, B.~S., {Moitinho}, A., \& {L{\'e}pine}, J.~R.~D. 2002,
  \aap, 389, 871

\bibitem[{{Dias} \& {L{\'e}pine}(2005)}]{2005ApJ...629..825D}
{Dias}, W.~S., \& {L{\'e}pine}, J.~R.~D. 2005, \apj, 629, 825

\bibitem[{{Dias} {et~al.}(2014){Dias}, {Monteiro}, {Caetano}, {L{\'e}pine},
  {Assafin}, \& {Oliveira}}]{2014A&A...564A..79D}
{Dias}, W.~S., {Monteiro}, H., {Caetano}, T.~C., {et~al.} 2014, \aap, 564, A79

\bibitem[{{Dias} {et~al.}(2019){Dias}, {Monteiro}, {L{\'e}pine}, \&
  {Barros}}]{2019MNRAS.486.5726D}
{Dias}, W.~S., {Monteiro}, H., {L{\'e}pine}, J.~R.~D., \& {Barros}, D.~A. 2019,
  \mnras, 486, 5726

\bibitem[{{Dias} {et~al.}(2021){Dias}, {Monteiro}, {Moitinho}, {L{\'e}pine},
  {Carraro}, {Paunzen}, {Alessi}, \& {Villela}}]{2021MNRAS.504..356D}
{Dias}, W.~S., {Monteiro}, H., {Moitinho}, A., {et~al.} 2021, \mnras, 504, 356

\bibitem[{{D{\'\i}az-Garc{\'\i}a} {et~al.}(2019){D{\'\i}az-Garc{\'\i}a},
  {Salo}, {Knapen}, \& {Herrera-Endoqui}}]{2019A&A...631A..94D}
{D{\'\i}az-Garc{\'\i}a}, S., {Salo}, H., {Knapen}, J.~H., \& {Herrera-Endoqui},
  M. 2019, \aap, 631, A94

\bibitem[{{Dobbs} {et~al.}(2022){Dobbs}, {Bending}, {Pettitt}, {Buckner}, \&
  {Bate}}]{2022MNRAS.517..675D}
{Dobbs}, C.~L., {Bending}, T.~J.~R., {Pettitt}, A.~R., {Buckner}, A.~S.~M., \&
  {Bate}, M.~R. 2022, \mnras, 517, 675

\bibitem[{{Dobbs} \& {Bonnell}(2006)}]{2006MNRAS.367..873D}
{Dobbs}, C.~L., \& {Bonnell}, I.~A. 2006, \mnras, 367, 873

\bibitem[{{Dobbs} \& {Pringle}(2010)}]{2010MNRAS.409..396D}
{Dobbs}, C.~L., \& {Pringle}, J.~E. 2010, \mnras, 409, 396

\bibitem[{{D'Onghia} {et~al.}(2013){D'Onghia}, {Vogelsberger}, \&
  {Hernquist}}]{2013ApJ...766...34D}
{D'Onghia}, E., {Vogelsberger}, M., \& {Hernquist}, L. 2013, \apj, 766, 34

\bibitem[{{Donor} {et~al.}(2018){Donor}, {Frinchaboy}, {Cunha}, \&
  et~al.}]{2018AJ....156..142D}
{Donor}, J., {Frinchaboy}, P.~M., {Cunha}, K., \& et~al. 2018, \aj, 156, 142

\bibitem[{{Donor} {et~al.}(2020){Donor}, {Frinchaboy}, \&
  et~al.}]{2020AJ....159..199D}
{Donor}, J., {Frinchaboy}, P.~M., \& et~al. 2020, \aj, 159, 199

\bibitem[{{Ferreira} {et~al.}(2020){Ferreira}, {Corradi}, {Maia}, {Angelo}, \&
  {Santos}}]{2020MNRAS.496.2021F}
{Ferreira}, F.~A., {Corradi}, W.~J.~B., {Maia}, F.~F.~S., {Angelo}, M.~S., \&
  {Santos}, J.~F.~C., J. 2020, \mnras, 496, 2021

\bibitem[{{Ferreira} {et~al.}(2019){Ferreira}, {Santos}, {Corradi}, {Maia}, \&
  {Angelo}}]{2019MNRAS.483.5508F}
{Ferreira}, F.~A., {Santos}, J.~F.~C., {Corradi}, W.~J.~B., {Maia}, F.~F.~S.,
  \& {Angelo}, M.~S. 2019, \mnras, 483, 5508

\bibitem[{{Gaia Collaboration} {et~al.}(2016){Gaia Collaboration}, {Brown}, \&
  {Vallenari}and~et al.}]{2016A&A...595A...2G}
{Gaia Collaboration}, {Brown}, A.~G.~A., \& {Vallenari}and~et al. 2016, \aap,
  595, A2

\bibitem[{{Gaia Collaboration} {et~al.}(2018){Gaia Collaboration}, {Brown},
  {Vallenari}, {Prusti}, {de Bruijne}, {Babusiaux}, {Bailer-Jones}, {Biermann},
  {Evans}, {Eyer}, {Jansen}, {Jordi}, {Klioner}, {Lammers}, {Lindegren},
  {Luri}, {Mignard}, {Panem}, {Pourbaix}, {Randich}, {Sartoretti}, {Siddiqui},
  {Soubiran}, {van Leeuwen}, {Walton}, {Arenou}, {Bastian}, {Cropper},
  {Drimmel}, {Katz}, {Lattanzi}, {Bakker}, {Cacciari}, {Casta{\~n}eda},
  {Chaoul}, {Cheek}, {De Angeli}, {Fabricius}, {Guerra}, {Holl}, {Masana},
  {Messineo}, {Mowlavi}, {Nienartowicz}, {Panuzzo}, {Portell}, {Riello},
  {Seabroke}, {Tanga}, {Th{\'e}venin}, {Gracia-Abril}, {Comoretto},
  {Garcia-Reinaldos}, {Teyssier}, {Altmann}, {Andrae}, {Audard},
  {Bellas-Velidis}, {Benson}, {Berthier}, {Blomme}, {Burgess}, {Busso},
  {Carry}, {Cellino}, {Clementini}, {Clotet}, {Creevey}, {Davidson}, {De
  Ridder}, {Delchambre}, {Dell'Oro}, {Ducourant},
  {Fern{\'a}ndez-Hern{\'a}ndez}, {Fouesneau}, {Fr{\'e}mat}, {Galluccio},
  {Garc{\'\i}a-Torres}, {Gonz{\'a}lez-N{\'u}{\~n}ez}, {Gonz{\'a}lez-Vidal},
  {Gosset}, {Guy}, {Halbwachs}, {Hambly}, {Harrison}, {Hern{\'a}ndez},
  {Hestroffer}, {Hodgkin}, {Hutton}, {Jasniewicz}, {Jean-Antoine-Piccolo},
  {Jordan}, {Korn}, {Krone-Martins}, {Lanzafame}, {Lebzelter}, {L{\"o}ffler},
  {Manteiga}, {Marrese}, {Mart{\'\i}n-Fleitas}, {Moitinho}, {Mora}, {Muinonen},
  {Osinde}, {Pancino}, {Pauwels}, {Petit}, {Recio-Blanco}, {Richards},
  {Rimoldini}, {Robin}, {Sarro}, {Siopis}, {Smith}, {Sozzetti}, {S{\"u}veges},
  {Torra}, {van Reeven}, {Abbas}, {Abreu Aramburu}, {Accart}, {Aerts},
  {Altavilla}, {{\'A}lvarez}, {Alvarez}, {Alves}, {Anderson}, {Andrei},
  {Anglada Varela}, {Antiche}, {Antoja}, {Arcay}, {Astraatmadja}, {Bach},
  {Baker}, {Balaguer-N{\'u}{\~n}ez}, {Balm}, {Barache}, {Barata}, {Barbato},
  {Barblan}, {Barklem}, {Barrado}, {Barros}, {Barstow}, {Bartholom{\'e}
  Mu{\~n}oz}, {Bassilana}, {Becciani}, {Bellazzini}, {Berihuete}, {Bertone},
  {Bianchi}, {Bienaym{\'e}}, {Blanco-Cuaresma}, {Boch}, {Boeche}, {Bombrun},
  {Borrachero}, {Bossini}, {Bouquillon}, {Bourda}, {Bragaglia}, {Bramante},
  {Breddels}, {Bressan}, {Brouillet}, {Br{\"u}semeister}, {Brugaletta},
  {Bucciarelli}, {Burlacu}, {Busonero}, {Butkevich}, {Buzzi}, {Caffau},
  {Cancelliere}, {Cannizzaro}, {Cantat-Gaudin}, {Carballo}, {Carlucci},
  {Carrasco}, {Casamiquela}, {Castellani}, {Castro-Ginard}, {Charlot},
  {Chemin}, {Chiavassa}, {Cocozza}, {Costigan}, {Cowell}, {Crifo}, {Crosta},
  {Crowley}, {Cuypers}, {Dafonte}, {Damerdji}, {Dapergolas}, {David}, {David},
  {de Laverny}, {De Luise}, {De March}, {de Martino}, {de Souza}, {de Torres},
  {Debosscher}, {del Pozo}, {Delbo}, {Delgado}, {Delgado}, {Di Matteo},
  {Diakite}, {Diener}, {Distefano}, {Dolding}, {Drazinos}, {Dur{\'a}n},
  {Edvardsson}, {Enke}, {Eriksson}, {Esquej}, {Eynard Bontemps}, {Fabre},
  {Fabrizio}, {Faigler}, {Falc{\~a}o}, {Farr{\`a}s Casas}, {Federici},
  {Fedorets}, {Fernique}, {Figueras}, {Filippi}, {Findeisen}, {Fonti},
  {Fraile}, {Fraser}, {Fr{\'e}zouls}, {Gai}, {Galleti}, {Garabato},
  {Garc{\'\i}a-Sedano}, {Garofalo}, {Garralda}, {Gavel}, {Gavras}, {Gerssen},
  {Geyer}, {Giacobbe}, {Gilmore}, {Girona}, {Giuffrida}, {Glass}, {Gomes},
  {Granvik}, {Gueguen}, {Guerrier}, {Guiraud}, {Guti{\'e}rrez-S{\'a}nchez},
  {Haigron}, {Hatzidimitriou}, {Hauser}, {Haywood}, {Heiter}, {Helmi}, {Heu},
  {Hilger}, {Hobbs}, {Hofmann}, {Holland}, {Huckle}, {Hypki}, {Icardi},
  {Jan{\ss}en}, {Jevardat de Fombelle}, {Jonker}, {Juh{\'a}sz}, {Julbe},
  {Karampelas}, {Kewley}, {Klar}, {Kochoska}, {Kohley}, {Kolenberg},
  {Kontizas}, {Kontizas}, {Koposov}, {Kordopatis}, {Kostrzewa-Rutkowska},
  {Koubsky}, {Lambert}, {Lanza}, {Lasne}, {Lavigne}, {Le Fustec}, {Le
  Poncin-Lafitte}, {Lebreton}, {Leccia}, {Leclerc}, {Lecoeur-Taibi},
  {Lenhardt}, {Leroux}, {Liao}, {Licata}, {Lindstr{\o}m}, {Lister}, {Livanou},
  {Lobel}, {L{\'o}pez}, {Managau}, {Mann}, {Mantelet}, {Marchal}, {Marchant},
  {Marconi}, {Marinoni}, {Marschalk{\'o}}, {Marshall}, {Martino}, {Marton},
  {Mary}, {Massari}, {Matijevi{\v{c}}}, {Mazeh}, {McMillan}, {Messina},
  {Michalik}, {Millar}, {Molina}, {Molinaro}, {Moln{\'a}r}, {Montegriffo},
  {Mor}, {Morbidelli}, {Morel}, {Morris}, {Mulone}, {Muraveva}, {Musella},
  {Nelemans}, {Nicastro}, {Noval}, {O'Mullane}, {Ord{\'e}novic},
  {Ord{\'o}{\~n}ez-Blanco}, {Osborne}, {Pagani}, {Pagano}, {Pailler},
  {Palacin}, {Palaversa}, {Panahi}, {Pawlak}, {Piersimoni}, {Pineau}, {Plachy},
  {Plum}, {Poggio}, {Poujoulet}, {Pr{\v{s}}a}, {Pulone}, {Racero}, {Ragaini},
  {Rambaux}, {Ramos-Lerate}, {Regibo}, {Reyl{\'e}}, {Riclet}, {Ripepi}, {Riva},
  {Rivard}, {Rixon}, {Roegiers}, {Roelens}, {Romero-G{\'o}mez}, {Rowell},
  {Royer}, {Ruiz-Dern}, {Sadowski}, {Sagrist{\`a} Sell{\'e}s}, {Sahlmann},
  {Salgado}, {Salguero}, {Sanna}, {Santana-Ros}, {Sarasso}, {Savietto},
  {Schultheis}, {Sciacca}, {Segol}, {Segovia}, {S{\'e}gransan}, {Shih},
  {Siltala}, {Silva}, {Smart}, {Smith}, {Solano}, {Solitro}, {Sordo}, {Soria
  Nieto}, {Souchay}, {Spagna}, {Spoto}, {Stampa}, {Steele},
  {Steidelm{\"u}ller}, {Stephenson}, {Stoev}, {Suess}, {Surdej}, {Szabados},
  {Szegedi-Elek}, {Tapiador}, {Taris}, {Tauran}, {Taylor}, {Teixeira},
  {Terrett}, {Teyssandier}, {Thuillot}, {Titarenko}, {Torra Clotet}, {Turon},
  {Ulla}, {Utrilla}, {Uzzi}, {Vaillant}, {Valentini}, {Valette}, {van Elteren},
  {Van Hemelryck}, {van Leeuwen}, {Vaschetto}, {Vecchiato}, {Veljanoski},
  {Viala}, {Vicente}, {Vogt}, {von Essen}, {Voss}, {Votruba}, {Voutsinas},
  {Walmsley}, {Weiler}, {Wertz}, {Wevers}, {Wyrzykowski}, {Yoldas},
  {{\v{Z}}erjal}, {Ziaeepour}, {Zorec}, {Zschocke}, {Zucker}, {Zurbach}, \&
  {Zwitter}}]{2018A&A...616A...1G}
{Gaia Collaboration}, {Brown}, A.~G.~A., {Vallenari}, A., {et~al.} 2018, \aap,
  616, A1

\bibitem[{{Gerhard}(2011)}]{2011MSAIS..18..185G}
{Gerhard}, O. 2011, Memorie della Societa Astronomica Italiana Supplementi, 18,
  185

\bibitem[{{Glushkova} {et~al.}(2010){Glushkova}, {Koposov}, {Zolotukhin},
  {Beletsky}, {Vlasov}, \& {Leonova}}]{2010AstL...36...75G}
{Glushkova}, E.~V., {Koposov}, S.~E., {Zolotukhin}, I.~Y., {et~al.} 2010,
  Astronomy Letters, 36, 75

\bibitem[{{Grand} {et~al.}(2012){Grand}, {Kawata}, \&
  {Cropper}}]{2012MNRAS.421.1529G}
{Grand}, R. J.~J., {Kawata}, D., \& {Cropper}, M. 2012, \mnras, 421, 1529

\bibitem[{{Griv} {et~al.}(2022){Griv}, {Gedalin}, \&
  {Jiang}}]{2022MNRAS.512.1169G}
{Griv}, E., {Gedalin}, M., \& {Jiang}, I.-G. 2022, \mnras, 512, 1169

\bibitem[{{Griv} {et~al.}(2021){Griv}, {Gedalin}, {Pietrukowicz}, {Majaess}, \&
  {Jiang}}]{2021MNRAS.502.4194G}
{Griv}, E., {Gedalin}, M., {Pietrukowicz}, P., {Majaess}, D., \& {Jiang}, I.-G.
  2021, \mnras, 502, 4194

\bibitem[{{Hao} {et~al.}(2021){Hao}, {Xu}, {Hou}, \&
  et~al.}]{2021AA...652A.102H}
{Hao}, C.~J., {Xu}, Y., {Hou}, L.~G., \& et~al. 2021, \aap, 652, A102

\bibitem[{{Hao} {et~al.}(2022{\natexlab{a}}){Hao}, {Xu}, {Wu}, {Lin}, {Bian},
  {Li}, \& {Liu}}]{2022AA...668A..13H}
{Hao}, C.~J., {Xu}, Y., {Wu}, Z.~Y., {et~al.} 2022{\natexlab{a}}, \aap, 668,
  A13

\bibitem[{{Hao} {et~al.}(2022{\natexlab{b}}){Hao}, {Xu}, {Wu}, {Lin}, {Liu}, \&
  {Li}}]{2022AA...660A...4H}
---. 2022{\natexlab{b}}, \aap, 660, A4

\bibitem[{{He} {et~al.}(2022{\natexlab{a}}){He}, {Li}, {Zhong}, \&
  et~al.}]{2022ApJS..260....8H}
{He}, Z., {Li}, C., {Zhong}, J., \& et~al. 2022{\natexlab{a}}, \apjs, 260, 8

\bibitem[{{He} {et~al.}(2022{\natexlab{b}}){He}, {Wang}, {Luo}, {Li}, {Liu}, \&
  {Jiang}}]{2022ApJS..262....7H}
{He}, Z., {Wang}, K., {Luo}, Y., {et~al.} 2022{\natexlab{b}}, \apjs, 262, 7

\bibitem[{{He} {et~al.}(2021){He}, {Xu}, {Hao}, {Wu}, \&
  {Li}}]{2021RAA....21...93H}
{He}, Z.-H., {Xu}, Y., {Hao}, C.-J., {Wu}, Z.-Y., \& {Li}, J.-J. 2021, Research
  in Astronomy and Astrophysics, 21, 093

\bibitem[{{Honig} \& {Reid}(2015)}]{2015ApJ...800...53H}
{Honig}, Z.~N., \& {Reid}, M.~J. 2015, \apj, 800, 53

\bibitem[{{Hou}(2021)}]{2021FrASS...8..103H}
{Hou}, L.~G. 2021, Frontiers in Astronomy and Space Sciences, 8, 103

\bibitem[{{Hou} \& {Han}(2014)}]{2014A&A...569A.125H}
{Hou}, L.~G., \& {Han}, J.~L. 2014, \aap, 569, A125

\bibitem[{{Hu} {et~al.}(2021){Hu}, {Zhang}, {Esamdin}, {Liu}, \&
  {Zeng}}]{2021ApJ...912....5H}
{Hu}, Q., {Zhang}, Y., {Esamdin}, A., {Liu}, J., \& {Zeng}, X. 2021, \apj, 912,
  5

\bibitem[{{Hunt} {et~al.}(2018){Hunt}, {Hong}, {Bovy}, {Kawata}, \&
  {Grand}}]{2018MNRAS.481.3794H}
{Hunt}, J. A.~S., {Hong}, J., {Bovy}, J., {Kawata}, D., \& {Grand}, R. J.~J.
  2018, \mnras, 481, 3794

\bibitem[{{Joshi}(2005)}]{2005MNRAS.362.1259J}
{Joshi}, Y.~C. 2005, \mnras, 362, 1259

\bibitem[{{Joshi}(2007)}]{2007MNRAS.378..768J}
---. 2007, \mnras, 378, 768

\bibitem[{{Joshi} {et~al.}(2016){Joshi}, {Dambis}, {Pandey}, \&
  {Joshi}}]{2016A&A...593A.116J}
{Joshi}, Y.~C., {Dambis}, A.~K., {Pandey}, A.~K., \& {Joshi}, S. 2016, \aap,
  593, A116

\bibitem[{{Karim} \& {Mamajek}(2017)}]{2017MNRAS.465..472K}
{Karim}, T., \& {Mamajek}, E.~E. 2017, \mnras, 465, 472

\bibitem[{{Kawata} {et~al.}(2014){Kawata}, {Hunt}, {Grand}, {Pasetto}, \&
  {Cropper}}]{2014MNRAS.443.2757K}
{Kawata}, D., {Hunt}, J. A.~S., {Grand}, R. J.~J., {Pasetto}, S., \& {Cropper},
  M. 2014, \mnras, 443, 2757

\bibitem[{{Kendall} {et~al.}(2011){Kendall}, {Kennicutt}, \&
  {Clarke}}]{2011MNRAS.414..538K}
{Kendall}, S., {Kennicutt}, R.~C., \& {Clarke}, C. 2011, \mnras, 414, 538

\bibitem[{{Kharchenko} {et~al.}(2013){Kharchenko}, {Piskunov}, {Schilbach},
  {R{\"o}ser}, \& {Scholz}}]{2013AA...558A..53K}
{Kharchenko}, N.~V., {Piskunov}, A.~E., {Schilbach}, E., {R{\"o}ser}, S., \&
  {Scholz}, R.~D. 2013, \aap, 558, A53

\bibitem[{{Kharchenko} {et~al.}(2007){Kharchenko}, {Scholz}, {Piskunov},
  {R{\"o}ser}, \& {Schilbach}}]{2007AN....328..889K}
{Kharchenko}, N.~V., {Scholz}, R.~D., {Piskunov}, A.~E., {R{\"o}ser}, S., \&
  {Schilbach}, E. 2007, Astronomische Nachrichten, 328, 889

\bibitem[{{Kounkel} {et~al.}(2020){Kounkel}, {Covey}, \&
  {Stassun}}]{2020AJ....160..279K}
{Kounkel}, M., {Covey}, K., \& {Stassun}, K.~G. 2020, \aj, 160, 279

\bibitem[{{Krumholz} {et~al.}(2019){Krumholz}, {McKee}, \&
  {Bland-Hawthorn}}]{2019ARA&A..57..227K}
{Krumholz}, M.~R., {McKee}, C.~F., \& {Bland-Hawthorn}, J. 2019, \araa, 57, 227

\bibitem[{{Lada} \& {Lada}(2003)}]{2003ARA&A..41...57L}
{Lada}, C.~J., \& {Lada}, E.~A. 2003, \araa, 41, 57

\bibitem[{{Liu} \& {Pang}(2019)}]{2019ApJS..245...32L}
{Liu}, L., \& {Pang}, X. 2019, \apjs, 245, 32

\bibitem[{{Loktin} \& {Beshenov}(2003)}]{2003ARep...47....6L}
{Loktin}, A.~V., \& {Beshenov}, G.~V. 2003, Astronomy Reports, 47, 6

\bibitem[{{Majaess} {et~al.}(2009){Majaess}, {Turner}, \&
  {Lane}}]{2009MNRAS.398..263M}
{Majaess}, D.~J., {Turner}, D.~G., \& {Lane}, D.~J. 2009, \mnras, 398, 263

\bibitem[{{Mermilliod} {et~al.}(2008){Mermilliod}, {Mayor}, \&
  {Udry}}]{2008AA...485..303M}
{Mermilliod}, J.~C., {Mayor}, M., \& {Udry}, S. 2008, \aap, 485, 303

\bibitem[{{Michikoshi} \& {Kokubo}(2018)}]{2018MNRAS.481..185M}
{Michikoshi}, S., \& {Kokubo}, E. 2018, \mnras, 481, 185

\bibitem[{{Minchev} {et~al.}(2011){Minchev}, {Famaey}, {Combes}, {Di Matteo},
  {Mouhcine}, \& {Wozniak}}]{2011A&A...527A.147M}
{Minchev}, I., {Famaey}, B., {Combes}, F., {et~al.} 2011, \aap, 527, A147

\bibitem[{{Monteiro} {et~al.}(2021){Monteiro}, {Barros}, {Dias}, \&
  {L{\'e}pine}}]{2021FrASS...8...62M}
{Monteiro}, H., {Barros}, D.~A., {Dias}, W.~S., \& {L{\'e}pine}, J. R.~D. 2021,
  Frontiers in Astronomy and Space Sciences, 8, 62

\bibitem[{{Myers} {et~al.}(2022){Myers}, {Donor}, \&
  et~al.}]{2022AJ....164...85M}
{Myers}, N., {Donor}, J., \& et~al. 2022, \aj, 164, 85

\bibitem[{{Netopil} {et~al.}(2022){Netopil}, {Oralhan}, {{\c{C}}akmak},
  {Michel}, \& {Karata{\c{s}}}}]{2022MNRAS.509..421N}
{Netopil}, M., {Oralhan}, {\.I}.~A., {{\c{C}}akmak}, H., {Michel}, R., \&
  {Karata{\c{s}}}, Y. 2022, \mnras, 509, 421

\bibitem[{{Netopil} {et~al.}(2016){Netopil}, {Paunzen}, {Heiter}, \&
  {Soubiran}}]{2016AA...585A.150N}
{Netopil}, M., {Paunzen}, E., {Heiter}, U., \& {Soubiran}, C. 2016, \aap, 585,
  A150

\bibitem[{{P{\'e}rez-Villegas} {et~al.}(2012){P{\'e}rez-Villegas}, {Pichardo},
  {Moreno}, {Peimbert}, \& {Vel{\'a}zquez}}]{2012ApJ...745L..14P}
{P{\'e}rez-Villegas}, A., {Pichardo}, B., {Moreno}, E., {Peimbert}, A., \&
  {Vel{\'a}zquez}, H.~M. 2012, \apjl, 745, L14

\bibitem[{{Poggio} {et~al.}(2021){Poggio}, {Drimmel}, {Cantat-Gaudin}, \&
  et~al.}]{2021A&A...651A.104P}
{Poggio}, E., {Drimmel}, R., {Cantat-Gaudin}, T., \& et~al. 2021, \aap, 651,
  A104

\bibitem[{{Poovelil} {et~al.}(2020){Poovelil}, {Zasowski}, \&
  et~al.}]{2020ApJ...903...55P}
{Poovelil}, V.~J., {Zasowski}, G., \& et~al. 2020, \apj, 903, 55

\bibitem[{{Portegies Zwart} {et~al.}(2010){Portegies Zwart}, {McMillan}, \&
  {Gieles}}]{2010ARA&A..48..431P}
{Portegies Zwart}, S.~F., {McMillan}, S. L.~W., \& {Gieles}, M. 2010, \araa,
  48, 431

\bibitem[{{Quillen} {et~al.}(2018){Quillen}, {Carrillo}, {Anders}, {McMillan},
  {Hilmi}, {Monari}, {Minchev}, {Chiappini}, {Khalatyan}, \&
  {Steinmetz}}]{2018MNRAS.480.3132Q}
{Quillen}, A.~C., {Carrillo}, I., {Anders}, F., {et~al.} 2018, \mnras, 480,
  3132

\bibitem[{{Reid} {et~al.}(2019){Reid}, {Menten}, \&
  et~al.}]{2019ApJ...885..131R}
{Reid}, M.~J., {Menten}, K.~M., \& et~al. 2019, \apj, 885, 131

\bibitem[{{Reid} {et~al.}(2014){Reid}, {Menten}, {Brunthaler}, {Zheng}, {Dame},
  {Xu}, {Wu}, {Zhang}, {Sanna}, {Sato}, {Hachisuka}, {Choi}, {Immer},
  {Moscadelli}, {Rygl}, \& {Bartkiewicz}}]{2014ApJ...783..130R}
{Reid}, M.~J., {Menten}, K.~M., {Brunthaler}, A., {et~al.} 2014, \apj, 783, 130

\bibitem[{{Roca-F{\`a}brega} {et~al.}(2013){Roca-F{\`a}brega}, {Valenzuela},
  {Figueras}, {Romero-G{\'o}mez}, {Vel{\'a}zquez}, {Antoja}, \&
  {Pichardo}}]{2013MNRAS.432.2878R}
{Roca-F{\`a}brega}, S., {Valenzuela}, O., {Figueras}, F., {et~al.} 2013,
  \mnras, 432, 2878

\bibitem[{{Ryu} \& {Lee}(2018)}]{2018ApJ...856..152R}
{Ryu}, J., \& {Lee}, M.~G. 2018, \apj, 856, 152

\bibitem[{{Schmeja} {et~al.}(2014){Schmeja}, {Kharchenko}, {Piskunov}, \&
  et~al.}]{2014AA...568A..51S}
{Schmeja}, S., {Kharchenko}, N.~V., {Piskunov}, A.~E., \& et~al. 2014, \aap,
  568, A51

\bibitem[{{Scholz} {et~al.}(2015){Scholz}, {Kharchenko}, {Piskunov},
  {R{\"o}ser}, \& {Schilbach}}]{2015AA...581A..39S}
{Scholz}, R.~D., {Kharchenko}, N.~V., {Piskunov}, A.~E., {R{\"o}ser}, S., \&
  {Schilbach}, E. 2015, \aap, 581, A39

\bibitem[{{Sellwood}(2011)}]{2011MNRAS.410.1637S}
{Sellwood}, J.~A. 2011, \mnras, 410, 1637

\bibitem[{{Siegert}(2019)}]{2019A&A...632L...1S}
{Siegert}, T. 2019, \aap, 632, L1

\bibitem[{{Sim} {et~al.}(2019){Sim}, {Lee}, {Ann}, \&
  {Kim}}]{2019JKAS...52..145S}
{Sim}, G., {Lee}, S.~H., {Ann}, H.~B., \& {Kim}, S. 2019, Journal of Korean
  Astronomical Society, 52, 145

\bibitem[{{Spina} {et~al.}(2021){Spina}, {Ting}, \&
  et~al.}]{2021MNRAS.503.3279S}
{Spina}, L., {Ting}, Y.~S., \& et~al. 2021, \mnras, 503, 3279

\bibitem[{{Tarricq} {et~al.}(2022){Tarricq}, {Soubiran}, {Casamiquela}, \&
  et~al.}]{2022AA...659A..59T}
{Tarricq}, Y., {Soubiran}, C., {Casamiquela}, L., \& et~al. 2022, \aap, 659,
  A59

\bibitem[{{Tarricq} {et~al.}(2021){Tarricq}, {Soubiran}, \&
  et~al.}]{2021AA...647A..19T}
{Tarricq}, Y., {Soubiran}, C., \& et~al. 2021, \aap, 647, A19

\bibitem[{{Taylor} \& {Cordes}(1993)}]{1993ApJ...411..674T}
{Taylor}, J.~H., \& {Cordes}, J.~M. 1993, \apj, 411, 674

\bibitem[{{Vande Putte} {et~al.}(2010){Vande Putte}, {Garnier}, {Ferreras},
  {Mignani}, \& {Cropper}}]{2010MNRAS.407.2109V}
{Vande Putte}, D., {Garnier}, T.~P., {Ferreras}, I., {Mignani}, R.~P., \&
  {Cropper}, M. 2010, \mnras, 407, 2109

\bibitem[{{Wada} {et~al.}(2011){Wada}, {Baba}, \&
  {Saitoh}}]{2011ApJ...735....1W}
{Wada}, K., {Baba}, J., \& {Saitoh}, T.~R. 2011, \apj, 735, 1

\bibitem[{{Xu} {et~al.}(2016){Xu}, {Reid}, {Dame}, \&
  et~al.}]{2016SciA....2E0878X}
{Xu}, Y., {Reid}, M., {Dame}, T., \& et~al. 2016, Science Advances, 2, e1600878

\bibitem[{{Yao} {et~al.}(2017){Yao}, {Manchester}, \&
  {Wang}}]{2017MNRAS.468.3289Y}
{Yao}, J.~M., {Manchester}, R.~N., \& {Wang}, N. 2017, \mnras, 468, 3289

\bibitem[{{Zhang} {et~al.}(2021){Zhang}, {Chen}, \&
  {Zhao}}]{2021ApJ...919...52Z}
{Zhang}, H., {Chen}, Y., \& {Zhao}, G. 2021, \apj, 919, 52

\bibitem[{{Zhong} {et~al.}(2022){Zhong}, {Chen}, {Jiang}, {Qin}, \&
  {Hou}}]{2022AJ....164...54Z}
{Zhong}, J., {Chen}, L., {Jiang}, Y., {Qin}, S., \& {Hou}, J. 2022, \aj, 164,
  54

\bibitem[{{Zhong} {et~al.}(2020){Zhong}, {Chen}, {Wu}, {Li}, {Bai}, \&
  {Hou}}]{2020AA...640A.127Z}
{Zhong}, J., {Chen}, L., {Wu}, D., {et~al.} 2020, \aap, 640, A127

\end{thebibliography}
\bibliographystyle{aasjournal}
\end{document}